\documentclass[a4paper,11pt]{article}
\pdfoutput=1 
\usepackage{jheppub2} 
\usepackage{bbm,bm,graphicx,mathtools,color,hyperref,slashed}
\usepackage{amssymb,amsmath}
\usepackage[T1]{fontenc}
\usepackage{comment}
\usepackage[all]{xy}
\usepackage[dvipsnames]{xcolor}
\usepackage{cancel}
 \usepackage{tikz}

\graphicspath{{./Figures/}}

\newcommand{\del}{\partial}

\newcommand{\WP}{W_{\text{P}}}
\newcommand{\WNP}{W_{\text{NP}}}
\preprint{.}

\title{Quantum geometry at arbitrary genus - I: Anharmonic potentials}
\author{Mustafa T\"{u}re,}
\author{Mithat \"{U}nsal}

\affiliation{Department of Physics and Astronomy, North Carolina State University, Raleigh, NC 27607, USA}

\emailAdd{mture@ncsu.edu, unsal.mithat@gmail.com}
\begin{abstract}
   {  
   In quantum mechanics of genus-1 potentials, 
 perturbation theory around the topologically trivial saddle and non-perturbative 
instanton saddle are constructively related via the P/NP relation.   Whether such a relation persists for higher-genus potentials has remained an open problem.  In this work, we resolve it using tools from complex algebraic geometry combined with exact WKB. The classical energy conservation relation $p^2 = 2(E-V(q))$ defines a genus-g Riemann surface $X$, whose periods are naturally organized by its Jacobian variety $J(X) = \mathbb{C}^{\text{g}}/\Lambda$ via the Abel-Jacobi map, and independently by the solution basis of the associated Picard-Fuchs equations. We show that the physically relevant, WKB-active cycles are generally related to the Picard-Fuchs basis by a linear transformation that lies in $\text{SL}(2\text{g},\mathbb{Z})$ but not in the symplectic group $\text{Sp}(2\text{g},\mathbb{Z})$ once $\text{g}\geq2$ — in contrast to the genus-1 case, where the two groups coincide. This mismatch is the obstruction that has made an explicit higher-genus P/NP relation elusive. We show that it can be resolved by passing to a modified Riemann bilinear identity, built from a suitably transformed intersection matrix, which the WKB-active periods do satisfy exactly. The result is an all-orders quantum P/NP relation valid for potentials of arbitrary genus, which we derive explicitly for genus-2 (quintic and sextic) and genus-3 (septic and octic) anharmonic oscillators.  }
\end{abstract}
\begin{document}
\maketitle
\section{Quantum Geometry of Exact-WKB Formalism}

In quantum mechanical systems  of genus-1 potentials, such as double-well potentials and periodic sine potential, there exists a remarkable relation that connects perturbative expansion around the trivial saddle to the one around the instanton \cite{Alvarez1, Alvarez3, Dunne:2013ada, Dunne:2014bca}. This is a constructive ``low order-low order"  relation\footnote{In contradistinction with standard resurgence, which leads to ``large order-  low order" relation, where the large order growth of the perturbation theory around the perturbative saddle has information about instanton action and small order perturbative fluctuations around the instanton-antiinstanton pair 
\cite{Brezin:1977gk,Bogomolny:1980ur,ZinnJustin:1981,ZinnJustin:1983,dillinger_resurgence_1993,zinn-justin_multi-instantons_2004-1,zinn-justin_multi-instantons_2004}; for reviews see \cite{Marino:2012zq,Dunne:2016nmc,Aniceto_2019,Dorigoni2019}}, which leads to the statement that if we know the perturbative expansion of energy as a function of level number $N$ and coupling $g$, $E(N, g)$ at some arbitrary order in $g$, we can easily derive the perturbative expansion around the instanton, 
 $e^{-S_I} P_{\rm fluc}(N, g)$ at the same order.  
 $E(N, g)$  can be calculated via Bender-Wu technique at any desired order \cite{Bender:1969si,Bender:1973rz, Sulejmanpasic:2016fwr}, while 
the perturbative expansion around an instanton using standard techniques, such as Feynman diagrams, is an extremely challenging and time-consuming task \cite{Escobar-Ruiz:2015nsa, Escobar-Ruiz:2015rfa,Dersy:2026jat,Dersy:2026ncf}.
The P/NP relation makes it conceptually clear that there is something much deeper about the structure of perturbation theory beyond what one can anticipate via Feynman diagrams. 
This deeper connection is through what one can call {\it quantum geometry}. It  has been explored for genus-1 potentials \cite{Basar:2015xna, Basar:2017hpr,Dunne:2016qix} and  connected to quasi-exact
solvability and to topological string theory in~\cite{Kozcaz:2016wvy,Codesido:2017jwp}. Its deformations and parametric versions
were studied in~\cite{Atakan1,Cavusoglu:2023bai}. 
The P/NP relation holds for genus-one potentials, but it fails in its original form for higher-genus potentials \cite{Gahramanov:2015yxk}. 
In this work, we aim to generalize this connection to higher-genus potentials. (See  also \cite{Codesido_2018, Fischbach_2019,  Misumi:2024gtf,Misumi:2025ijd, Kreshchuk_2019,Kreshchuk_2019_2, Meynig:2024nam, Gulden_2015, Raman:2020sgw, Alim:2022oll, DelMonte:2022kxh}). We outline  an algorithm  for  key ingredients below.  The first three items are on how to connect classical mechanics to Riemann surfaces and complex algebraic geometry (see \cite{griffiths1989introduction} for an introduction to complex algebraic curves), and the last three items are essentially quantum geometry and   asymptotic analysis in $\hbar$. 

\begin{itemize} 
\item  {\bf Riemann surface, $X$: } View classical energy conservation relation  
 \begin{align} 
 p^2 (q) = 2(u - V(q))   
 \label{classical}
 \end{align}
as a complex algebraic curve (a Riemann surface)  $X$.   In classical mechanics, the motion is only executed in regimes where $V(q) \leq  u$, the classically allowed regions. Hence, 
it suffices to describe  classical mechanics in terms of real algebraic curves in phase space.   In contrast, in quantum mechanics, classically forbidden regions are also allowed due to tunneling.  Therefore, the complex algebraic curves  are for quantum mechanics what real algebraic curves  (or phase space curves) are  for  classical mechanics. 

\item  {\bf Classical Periods of $X$: }  The (classical) periods  of the genus-g curve  $X$ plays a prominent role in the quantization.      Let $\rm g$ denote the genus of the curve $X$, then  $X$ has 2g incontractible and homotopically distinct loops $\left\{\mathbf{a}_i,\mathbf{b}_i\right\}_{i=1}^{\text{g}}\subset H_1(X,\mathbb{Z})$ which form  a canonical basis for the first homology group $H_1(X,\mathbb{Z})$.
  Let $2\text{g}$ complex abelian differentials $\left\{\omega_k=\frac{q^{k-1}dq}{y}\right\}_{k=1}^{\rm 2g}$  be a basis of differential one-forms of the cohomology group\footnote{More specifically we'll be working with the de Rham cohomology $H^1_{\text{dR}}(X,\mathbb{C})$.} $H^1(X,\mathbb{C})$.  
     The integral of the $2\text{g}$-differentials over the basis of the homology cycles defines what is called the period \footnote{For this reason each Abelian differential of the first kind $\omega_k$ for $1\leq k\leq \text{g}$ will be referred to as \text{period differentials} and the unique Abelian differential $\omega_0\equiv ydq$ will be referred to as the \text{action differential.}}
     \begin{equation}
         \tau_{ij} = \oint_{\gamma_i}\omega_j.
     \end{equation}

\item  {\bf Jacobian variety,  $J(X)$:}     In practice, it is difficult to work with a  Riemann surface. 
   An extremely valuable tool to study the properties of a  Riemann surface $X$ is through its Jacobian variety, $J(X)$.   
For a potential corresponding to a genus-g surface, the  Jacobian variety  takes the geometric form of a complex torus, 
\begin{align}
J(X) = \mathbb{C}^\text{g} / \Lambda 
\label{JV}
\end{align}
where the genus of the curve dictates the dimension of the torus, and $\Lambda$ is the rank-g period lattice.  The Jacobian variety has a natural group structure that makes it more suitable to work with. 

\item  {\bf Abel-Jacobi map:}   The Abel-Jacobi map provides a concrete way to get from the Riemann surface to its Jacobian, connecting geometry to algebraic structure.  Choosing a pair of fixed starting and arbitrary destination points $(P_0, P)$ on $X$,  we can integrate the 
complex abelian differentials $\left\{\omega_k\right\}_{k=1}^{\rm 2g}$  along a path, up to an element of $\Lambda$. This gives us a complex vector with g 
components as an element of its equivalence class on the Jacobian $J(X)$. 
    \begin{align}
    \mu(P) = \left( \int_{P_0}^{P} \omega_1, \ \int_{P_0}^{P} \omega_2, \ \dots, \ \int_{P_0}^{P} \omega_g \right) \in\mathbb{C}^\text{g}/\Lambda
    \end{align}

\item  {\bf Exact WKB and quantum periods: } The Schr\"odinger equation can be recast into the non-linear Riccati equation (quantum Hamilton-Jacobi equation), which can be solved recursively, yielding asymptotic series. (The roots of exact WKB are in historical works \cite{Voros:1983, Silverstone:1985, Delabaere:1997,Delabaere:1997pham,dillinger_resurgence_1993,DP1,Kawai1}.)
These solutions can be used to formally represent quantum periods. A quantum period is a formal asymptotic expansion that can be written as
\begin{equation}
    W_{\gamma}(u,\hbar) = \frac{1}{\hbar}\left(\sum_{n=0}^{\infty}\hbar^n\mathcal{D}_n\right)    W_{\gamma}^{(0)}(u), \qquad  W_{\gamma}^{(0)}(u) = \oint_{\gamma}pdq,  
    \label{QuantumPeriod}
    \end{equation}
where  $\mathcal{D}_n$ are the \textit{quantum generators} which are linear differential operators in $u$, and  $W_{\gamma}^{(0)}(u)$ are the \textit{classical actions}. 

\item  {\bf Picard-Fuchs equations:} Construct  the Picard-Fuchs equations satisfied by the classical action, which are ordinary linear differential equations of order (at most) 2g.  
This can be done by using the completeness of the cohomology basis. The procedure is given in the appendix. 
PF-equations have $2 {\rm g}$ linearly independent solutions. These solutions  also define a canonical basis (which we refer to as  \textit{PF-basis}) in correspondence with  the classical actions   $W_{\gamma}^{(0)}(u)$. 

The action of quantum generators $\mathcal{D}_n$ on the classical periods as in \eqref{QuantumPeriod}  constructs the formal power series expansion for quantum periods.  \footnote{ Quantum periods  also appear in the context of 
 $\mathcal{N}=2$ gauge theories and
refined topological strings in the Nekrasov--Shatashvili
limit~\cite{Nekrasov:2009rc,Mironov:2009uv,Aganagic:2011mi,Huang:2011qx,Kashani-Poor:2015pca,Ashok:2016yxz,Grassi:2019coc, Klemm:1995wp,MATONE1995342,Ito:2017iba,Ito:2018hwp}.  The resurgent structure of quantum periods was analyzed
in~\cite{Gu:2022fss,Ito:2018eon,Imaizumi:2020fxf,Emery:2020qqu,Ito:2024odeim}. }

\item{\bf WKB active and passive cycles:} Consider the exact WKB analysis for a general genus-g surface.  Depending on the moduli parameters of the potentials and energy, there is a set of active and passive cycles. 
Active (or physical) cycles contribute to the quantization condition.  The active cycles can be written as a linear combination of the canonical PF-basis. The passive linear combinations are the complement, and do not enter the exact quantization condition. 
Which cycles are active and which are passive depends on the position in the moduli space. 

 \item{\bf Higher genus P/NP relation:}  
 The physical cycles that enters into the spectrum (WKB analysis) as a linear combinations of 
 Picard-Fuchs basis.   Then,  the quantum P/NP relation can naturally be viewed as the quantum version of a \textit{modified} Riemann bilinear identity.  
 We provide a concrete example for a set of genus-2 and genus-3 potentials. 

\end{itemize}

   

The general formalism developed in Sec.~\ref{sec:General} and
Sec.~\ref{sec:PNP} applies to arbitrary classical multi-well
potentials~\eqref{classical}. In Sec.~\ref{sec:examples}, however, we
restrict our applications to genus-2 and genus-3 systems with a single
instanton type, rather than genuinely multi-well systems supporting
several distinct instantons. This restriction is a matter of
convenience rather than principle: potentials of this class are simply
more tractable higher-genus examples to work with explicitly.


\section{Quantum Actions and Periods of a System}
\label{sec:General}
\begin{figure}[htb]
    \centering
    \includegraphics[width=1\linewidth]{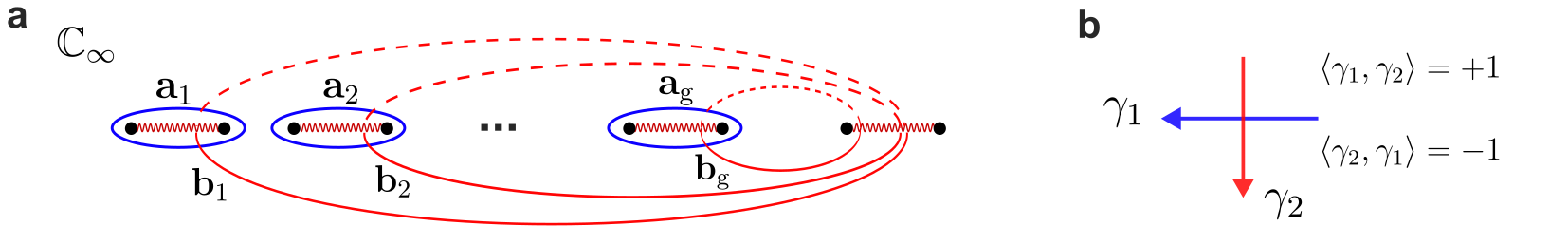}
    \caption{\textbf{Canonical basis cycles and intersection numbers.} \textbf{a,} A choice for a set of canonical basis cycles $\{(\mathbf{a}_i, \mathbf{b}_i)\}_{i=1}^{\text{g}}$ of a Riemann surface of genus g. The squiggly dark red lines indicate the branch cuts emanating from pairs of branch points, and the dashed segments of cycles indicate they lie on the second sheet. This basis set is equivalent to any other choice of canonical basis set up to an $\text{Sp}(2\text{g},\mathbb{Z})$ transformation. \textbf{b,} The intersection number structure of two cycles $\gamma_1$ and $\gamma_2$. The homology space is endowed with the inner product shown, which gives the space its symplectic structure defined by the intersection matrix \eqref{intersection}.} 
    \label{fig:1}
\end{figure}

\begin{figure}[htb]
    \centering
    \includegraphics[width=1\linewidth]{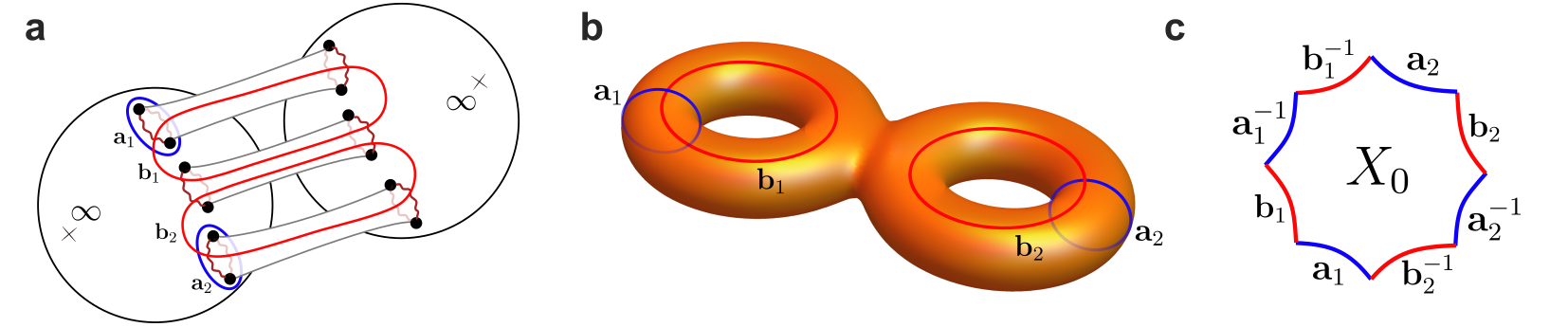}
    \caption{\textbf{A genus 2 Riemann surface $p^2 = f(q)$ and its geometric and topological representations.} \textbf{a,} The gluing of two compactified complex planes $\mathbb{C}_{\infty}$ from their branch cuts which gives the compact Riemann surface $X$. A set of canonical basis cycles $\{(\textbf{a}_i,\textbf{b}_i)\}_{i=1}^{2}$ is represented on it by the blue and red loops. The squiggly dark red lines represent the branch cuts where the two surfaces are glued. \textbf{b,} A torus, $T^{\text{2}}$, of genus 2 which is homeomorphic to the Riemann surface X. Same canonical basis cycles are shown as in $\textbf{a}$. \textbf{c,} The canonical dissection of the Riemann surface $X$ using the canonical basis pairs. $X_0$ is the interior of the Riemann surface and The cycles are the boundary $\del X$.}
    \label{fig:2}
\end{figure}

For completeness, let us quickly introduce the exact WKB framework \cite{DP1, Kawai1, Ture:2024qhj, Iwaki1, Gaiotto:2009hg,  vanSpaendonck:2023znn,Bucciotti:2023trp, sueishi_exact-wkb_2020, sueishi_exact-wkb_2021, Fischbach_2019, nikolaev2024geometry}. The Schr\"odinger equation is transformed into the quantum Hamilton-Jacobi equation by using the quantum action ansatz $\psi = Ae^{\frac{i}{\hbar}S_Q}$ 
\begin{equation}
    -\frac{\del S_Q}{\del t} = \frac{1}{2}\left(\frac{\del S_Q}{\del q}\right)^2 + V(q) - \frac{i\hbar}{2}\frac{\del ^2 S_Q}{\del q^2}
\end{equation}
where $S_Q$ is the quantum Hamilton's principle function $S_Q=-ut + \int^qp_Q(q',u,\hbar)dq'$. This results in an equation for the quantum momentum function $p_Q$,
\begin{equation}
    p_Q^2 = 2\left(u-V(q)\right) + i\hbar\frac{\del p_Q}{\del q}.
\end{equation}
This equation can be solved by a formal asymptotic expansion in $\hbar$, giving the following recursion relation for the quantum action differential's coefficients $p_Q dq = \sum_{n=0}^{\infty}\frac{\rho_n}{p^{3n-1}}dq\hbar^n$ with $\rho_n$ satisfying
\begin{equation}
    \rho_{n+1}(q,u) = \frac{i}{2}\left(p^2\frac{\del\rho_{n}}{\del q}+(3n-1)\frac{dV}{dq}\rho_n\right)-\frac{1}{2}\sum_{k=1}^n\rho_k\rho_{n+1-k}
\end{equation}
where $\rho_0=0$, $\rho_1=- \frac{i}{2} V^{'}$, and $p^2 = 2(u-V)$. This construction allows us to define the quantum reduced actions for each periodic classical orbit in the phase space.
The reduced quantum actions are defined as
\begin{align}
    a_i(u,\hbar) =\frac{1}{2\pi \hbar} \sum_{n=0}^{\infty}\hbar^{n}\oint_{\mathbf{a}_i}\frac{\rho_{n}(q,u)}{p^{3n-1}}dq,\qquad a_i^D(u,\hbar)=\frac{-i}{\hbar}\sum_{n=0}^{\infty}\hbar^{n}\oint_{\mathbf{b}_i}\frac{\rho_{n}(q,u)}{p^{3n-1}}dq
\end{align}
respectively, where each $\frac{\rho_n}{p^{3n-1}}dq\in H^1(X,\mathbb{C})$. 
We assume that $V(q)$ is some potential that can be represented as a polynomial in $q$ with real coefficients and exhibit at least one harmonic minimum. Therefore, $X$ is the Riemann surface defined by the (hyper)elliptic curve $\mathcal{E}$ \footnote{The form of the curve is same as  the    Seiberg-Witten curve for $SU(N)$  gauge theory with genus $\text{g}= N-1$ \cite{Seiberg:1994rs,Klemm:1995wp}.}
\begin{equation}\label{curve}
    \mathcal{E} :p^2 =2\left(u-V(q)\right)
\end{equation}
of genus $\text{g}=\lfloor\frac{\text{deg}V(q)-1}{2}\rfloor$, and the set of one cycles $\left\{\mathbf{a}_i,\mathbf{b}_i\right\}_{i=1}^{\text{g}}\subset H_1(X,\mathbb{Z})$ define a canonical basis for the first homology group $H_1(X,\mathbb{Z})$ satisfying the intersection properties, shown in Fig.\textbf{\ref{fig:1}},
$$\langle\mathbf{a}_i,\mathbf{a}_j\rangle =0,\;\langle\mathbf{b}_i,\mathbf{b}_j\rangle =0,\;\langle\mathbf{a}_i,\mathbf{b}_j\rangle =-\langle\mathbf{b}_j,\mathbf{a}_i\rangle=\delta_{ij}, \quad \forall i,j\in\{1,2,\cdots,\text{g}\}.$$
Canonical basis cycles and their intersection properties defines a symplectic structure on the Riemann surface, whose symplectic intersection matrix $J$ is defined by
\begin{equation}\label{intersection}
   J\coloneq \begin{pmatrix}
        0 & I_{\text{g}}\\
        -I_{\text{g}} & 0
    \end{pmatrix} \equiv \begin{pmatrix}
        \langle\vec{\mathbf{a}},\vec{\mathbf{a}}\rangle & \langle\vec{\mathbf{a}},\vec{\mathbf{b}}\rangle\\
        \langle\vec{\mathbf{b}},\vec{\mathbf{a}}\rangle & \langle\vec{\mathbf{b}},\vec{\mathbf{b}}\rangle
    \end{pmatrix}
\end{equation}
where $I_{\text{g}}$ is the $\text{g}\times\text{g}$ identity matrix.
Moreover, the cohomology group $H^1(X,\mathbb{C})$ is spanned by $2\text{g}$ complex abelian differentials, $\left\{\omega_i\right\}_{i=1}^{2\text{g}}$. If one chooses $\omega_1=p_Q(q,u,\hbar)dq$, and $\omega_2=\frac{\del}{\del u}p_Q(q,u,\hbar)dq$ then the pairs of quantum action and period integrals are just projections on the two coordinates of the Abel-Jacobi map $A_u(P)$ defined as a function from the Riemann surface $X$ to its Jacobi variety $\mathbb{C}^{\text{g}}/\Lambda\simeq T^{\text{g}}$ (see Figure \textbf{\ref{fig:2}a,b}),
\begin{equation}
    A_u:P \mapsto \left(\int_{P_0}^P\omega_1,\;\int_{P_0}^P\omega_2,\;\cdots,\;\int_{P_0}^P\omega_{\text{g}}\right)\; \text{mod}\,\Lambda
\end{equation}
for any path from $P_0$ to $P$, where $\Lambda=\Lambda_{\omega_1}\times\Lambda_{\omega_2}\times\cdots\times\Lambda_{\omega_g}$ is the rank-g quantum period lattice and
\begin{equation}
    \Lambda_{\omega_k} \coloneq \left\{\sum_{n,m\in\mathbb{Z}}\sum_{i=1}^{\text{g}}\left(n\oint_{\textbf{a}_i}\omega_k+m\oint_{\textbf{b}_i}\omega_k\right)\;\Big|\; \omega_k\in\bigoplus_{n=0}^{\infty}\hbar^{n} H^1(X,\mathbb{C})\right\}.
\end{equation}
The choice of cycles for the canonical basis pairs are up to the problem at hand. One can choose the cycles so that each cycle encircles the $2\text{g}+2$ turning points defined by the curve $p=0$ for nonsingular modulus $u$ and such that the set of cycles forms a canonical dissection of $X$, labeled as $X_0$, as shown in Fig.\textbf{\ref{fig:2}c}. However, in what follows we will show how a useful choice can be made naturally and how they are related to the relevant physical one-cycles.

\subsection{Picard-Fuchs Equations and Quantum Generators}
Because the cohomology group $H^1(X,\mathbb{C})$ has $2\text{g}$ real dimensions, any linear combination of $2\text{g}+1$ real Abelian differentials on $X$ will be linearly dependent up to an exact form. We can generate $2\text{g}$-many differentials by taking a derivative of the classical one-form $pdq$ with respect to the modulus $u$. Hence,
\begin{equation}\label{PF}
\sum_{k=0}^{2\text{g}}f_k^{(0)}(u)\frac{\del^k}{\del u^k}pdq = dF,
\end{equation}
where $f_k^{(0)}(u)\in \mathbb{Z}[u]$ are polynomials of the modulus $u$ with integer coefficients and $dF$ is an exact form in $H^1(X,\mathbb{C})$. The highest order coefficient $f^{(0)}_{2\text{g}}(u)$ is always proportional to the discriminant $\Delta_q(\mathcal{E})=\prod_{i=1}^{d-1}(u-V(q_i^{(c)}))$ of the hyperelliptic curve $\mathcal{E}$, where $q_i^{c}$ are the saddles of $V(q)$. Upon integration of $\eqref{PF}$ over any nontrivial 1-cycle, we get the so-called Picard-Fuchs equation (see \cite{Griffiths:1969a, Griffiths:1969b, Fischbach_2019, Kreshchuk_2019, Kreshchuk_2019_2, Gulden_2015})
\begin{equation}\label{PFequation}
    \mathcal{L}_{PF}\oint_{\gamma}p(u)dq \coloneq \sum_{n=0}^{2\text{g}}f_k^{(0)}(u)\frac{\del^k}{\del u^k}\oint_{\gamma}p(u)dq=0.
\end{equation}
Similarly, since each order in the quantum action differential $p_{Q,n}dq = \frac{\rho_n}{p^{3n-1}}dq$ is an element of the same cohomology group, there exists a linear relationship between each of them and the classical action differential. Again, up to an exact form,
\begin{equation}
    \frac{\rho_n}{p^{3n-1}}dq -\sum_{k\in P_n[g]}h^{(n)}_k(u)\frac{\del^{k}}{\del u^{k}}pdq=dF
\end{equation}
where $P_n[\text{g}]=\left\{(n-2)\text{g}, \cdots, n\text{g}-1\right\}$ is the $n^{\text{th}}$ element of the $2\text{g}$ partitioning of natural numbers with zero included and $h_k(u)\in\mathbb{Q}[u]$ is a polynomial of $u$ with rational coefficients. After integrating over a nontrivial cycle, we find the generating relation of quantum corrections to the classical action
\begin{equation}
    \mathcal{D}_n\oint_{\gamma}pdq = \oint_{\gamma}dq\frac{\rho_n}{p^{3n-1}}
\end{equation}
where we identified the \textit{quantum generators}, $\mathcal{D}_n$, to be
\begin{equation}
    \mathcal{D}_n \coloneq \sum_{k\in P_n[g]}h^{(n)}_k(u)\frac{\del^{k}}{\del u^{k}}
\end{equation}
which are linear differential operators acting on the classical action. Therefore, one can write any quantum action as a tower of quantum generators acting on the classical reduced action, which is a linear combination of the solution to the Picard-Fuchs equation \eqref{PF},
\begin{equation}\label{quantumaction}
    a_{\gamma}(u,\hbar) = \frac{1}{\hbar}\left(\sum_{n=0}^{\infty}\hbar^n\mathcal{D}_n\right)\oint_{\gamma}pdq, \qquad \gamma=\sum_{i=1}^{\text{g}}n_i\mathbf{a}_i+m_i\mathbf{b}_i,\qquad n_i,m_i\in\mathbb{Z}.
\end{equation}
The representations of quantum generator operators are not unique; one can fix the order of derivatives for all n, in exchange for allowing $h_k^{(n)}(u)$ to be rational functions of $u$.

The existence of the quantum generators shows that the quantum period lattice $\Lambda_Q$ is a straightforward extension of the classical period lattice $\Lambda_{pdq}$ as an infinite tower of lattices. 
\subsection{The Modular and Symplectic Group Actions on Canonical Basis}
Let $$\text{PF}_{\text{classical}}=\left\{\left(a_i^{(0)}(u),a_{i,D}^{(0)}(u)\right)\right\}_{i=1}^{\text{g}}$$ be the set of linearly independent general solutions to the Picard-Fuchs equation \eqref{PFequation}. We can use these general solutions to construct canonical action pairs by acting on them with the quantum action generators,
\begin{equation}
\text{PF}=\left\{\big(a_i(u,g),a_{i,D}(u,g) \big)=\frac{1}{g}\left(\sum_{n=0}^{\infty}g^n\mathcal{D}_n\right)\big(a_i^{(0)}(u),a_{i,D}^{(0)}(u)\big)\Bigg|i=1,2,\cdots,\text{g} \right\}
\end{equation} 
where the small perturbative expansion parameter $g$ is on the same footing with $\hbar$ as written in \eqref{quantumaction}. The set of actions, PF, spans the quantum action lattice $\Lambda_{p_Qdq}$. However, we still do not have any access to their topological information. Therefore, we can not be certain that their respective integration cycles form pairs of canonical cycles. We will refer to their associated cycles $(\vec{\textbf{a}},\vec{\textbf{b}})$ as \textit{PF cycles}, and the corresponding actions $(\vec{a},\vec{a}_D)$ as \textit{PF basis}. Here, we have introduced the notation $(\vec{x},\vec{y})\coloneq(x_1,x_2,\cdots,x_{\text{g}},y_1,y_2,\cdots,y_\text{g})$ or also in a column vector form,
\begin{equation}
    \begin{pmatrix}
        \vec{a}\\
        \vec{a}_D
    \end{pmatrix}\coloneq\begin{pmatrix}
        a_1\\
        a_2\\        \vdots\\
        a_\text{g}\\
        a_1^D\\
        \vdots\\
        a_{\text{g}}^D
    \end{pmatrix} \qquad \text{and}\qquad 
    \begin{pmatrix}
        \vec{\textbf{a}}\\
        \vec{\textbf{b}}
    \end{pmatrix}\coloneq\begin{pmatrix}
        \textbf{a}_1\\
        \textbf{a}_2\\
        \vdots\\
        \textbf{a}_\text{g}\\
        \textbf{b}_1\\
        \vdots\\
        \textbf{b}_{\text{g}}
    \end{pmatrix} 
\end{equation}
For notational ease, we will also define the period vectors of a given one-form $\omega$ as
\begin{equation}
    \Pi_{\omega} \coloneq \begin{pmatrix}
        \vec{a}_{\omega}\\
        \vec{a}^D_{\omega}
    \end{pmatrix}.
\end{equation}
Notice the PF-basis actions are still derived from the set of linearly independent \textbf{general solutions} so they correspond to a pair of quantum actions up to some real scaling factor. Therefore, for a given specific system we still need to scale each of the PF-basis actions with some diagonal linear matrix in order to fix the quantum one form they correspond to. Let
\begin{equation}
    S=\text{diag}(c_1,\cdots,c_\text{g},c_1^D,\cdots,c_{\text{g}}^D), \quad c_i,c_i^D\in\mathbb{R}
\end{equation}
be the $2\text{g}\times2\text{g}$ diagonal scaling matrix for the specific set of PF-basis actions. The coefficients can be calculated by direct integration of the classical action up to $2\text{g}$ orders in the moduli parameter $u$ along a nontrivial cycle. If S has the block diagonal form
\begin{equation}
    S = \begin{pmatrix}
        C & 0 \\
        0 & (C^t)^{-1}
    \end{pmatrix}
\end{equation}
then $S\in \text{Sp}(2\text{g},\mathbb{R})$ which respects the intersection structure of $X$ as an extension of the $\text{Sp}(2\text{g},\mathbb{Z})$ symmetry satisfied by the canonical basis cycles. Otherwise, the symplectic intersection matrix $J$ needs to be modified, and one must define a new symplectic structure.

Indeed, this is usually the case for the PF-basis. Since their cycles do not necessarily form a canonical pair, one needs to change the rule of intersection by defining a new intersection matrix $K$ that the PF-cycle pairs satisfy
\begin{equation}\label{PFintersection}
    K\coloneq \begin{pmatrix}
        0& \kappa_{\text{g}}\\
        -\kappa_{\text{g}}& 0
    \end{pmatrix} = \begin{pmatrix}
        \langle \vec{\textbf{a}},\vec{\textbf{a}}\rangle & \langle\vec{\textbf{a}},\vec{\textbf{b}\rangle}\\
        \langle\vec{\textbf{b}},\vec{\textbf{a}}\rangle&\langle\vec{\textbf{b}},\vec{\textbf{b}}\rangle
    \end{pmatrix}
\end{equation}
where $\kappa_\text{g}$ is a diagonal $\text{g}\times\text{g}$ matrix.
 For a given genus-g Riemann surface $X$ defined by the curve $\mathcal{E}$ in \eqref{curve}, another set of pairs of cycles $(\vec{\textbf{a}'},\vec{\textbf{b}'})\subset H_1(X,\mathbb{Z})$ respects the same intersection matrix \textit{if and only if} the corresponding actions can be written as a linear transformation of the PF basis
\begin{equation}
    \begin{pmatrix}
        \vec{a}'\\
        \vec{a}'_D
    \end{pmatrix} = M\begin{pmatrix}
        \vec{a}\\
        \vec{a}_D
    \end{pmatrix} 
\end{equation}
given that the $2\text{g}\times 2\text{g}$ matrix $M$ is an element of the \textit{symplectic group}, i.e., $M\in \text{Sp}(2\text{g},\mathbb{Z})$ or equivalently
\begin{equation}
    M^{T}KM = K
\end{equation}
with $K$ being the new symplectic intersection matrix defined in \eqref{PFintersection}. For $\text{g}=1$, the symplectic group and the modular group are exactly the same, $\text{Sp}(2,\mathbb{Z}) = \text{SL}(2,\mathbb{Z})$. However, for $\text{g}\geq 2$, the symplectic group is a strict subgroup of the special linear group $\text{Sp}(2\text{g},\mathbb{Z})\subset \text{SL}(2\text{g},\mathbb{Z})$. This is the key observation for understanding the P/NP relation in higher-genus systems.

\begin{figure}[htb]
    \centering
    \includegraphics[width=1\linewidth]{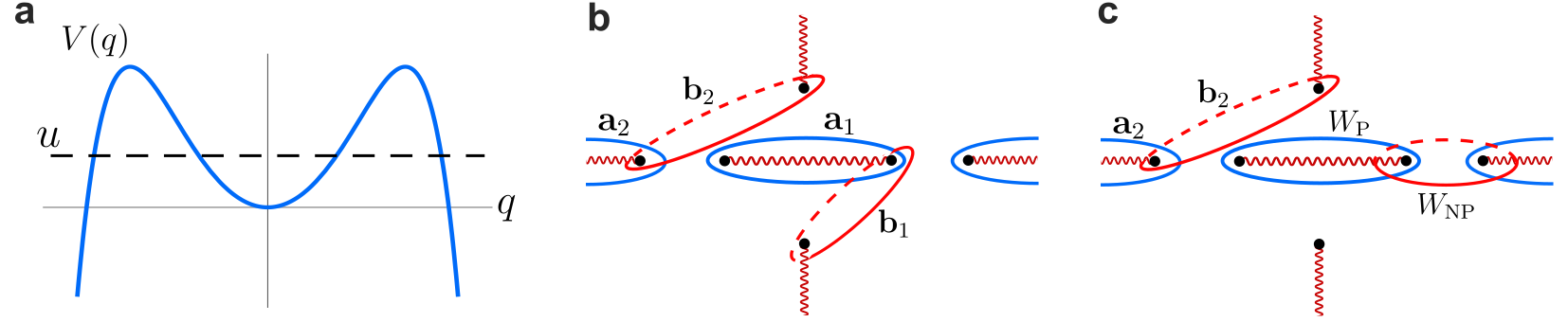}
    \caption{\textbf{A genus 2 sextic anharmonic oscillator  potential $V(q) = \frac{1}{2}q^2-q^6$ and its branch cut structure.} \textbf{a,} The profile of the potential on the real axis shown in blue and a nonsingular value of the modulus (energy) parameter $u$ shown by the dashed horizontal line. \textbf{b,} The set of canonical basis cycles represented by an element of the equivalence class of canonical basis cycles under the action of $\text{Sp}(2\text{g},\mathbb{Z})$. \textbf{c,} WKB-active cycles that participate in the quantization condition denoted by $(W_{\text{P}}, W_{\text{NP}})$. The complete set of cycles is related to the canonical cycles by some linear transformation $M_o(\mathbf{a}_1,\mathbf{a}_2,\mathbf{b}_1, \mathbf{b}_2)^T = (W_{\text{P}}, \mathbf{a}_2, W_{\text{NP}},\mathbf{b}_2)^T$, where $M_o\in \text{SL}(2\text{g},\mathbb{Z})$. Clearly, $W_{\text{NP}}$ does not respect the intersection rules for the canonical basis pairs.}
    \label{fig:3}
\end{figure}

Now we argue that for a given system defined by the classical curve $\mathcal{E}$, there exists a product element $M_*=M_oS\in \text{GL}(2\text{g},\mathbb{R})$ with $M_o\in \text{SL}(2\text{g},\mathbb{Z})$ and $S=\text{diag}(c_i)\in \text{GL}(2g,\mathbb{R})$ such that its row space defines the linear combinations for the \textit{WKB-active} perturbative and nonperturbative actions $(\vec{W},\vec{W}_D)$ and their one-cycles $(\vec{\gamma},\vec{\gamma}_D)$ in terms of the PF-basis actions and PF-basis one-cycles,
\begin{align}\label{transform}
    \begin{pmatrix}
        \vec{W}\\
        \vec{W}_D
    \end{pmatrix} = M_*\begin{pmatrix}
        \vec{a}\\
        \vec{a}_D
    \end{pmatrix} \quad \text{and} \quad  \begin{pmatrix}
        \vec{\gamma}\\
        \vec{\gamma}_D
    \end{pmatrix} = M_o\begin{pmatrix}
        \vec{\textbf{a}}\\
        \vec{\textbf{b}}
    \end{pmatrix}
\end{align}
Since $M_o\in\text{SL}(2\text{g},\mathbb{Z})$, its inverse $M_o^{-1}$ also exists in the same modular group. It is this specific transformation $M_o$ and the corresponding linear combinations of each element that will be used in our formulation of the P/NP relation. Our goal is to compute the new intersection structure of the PF-basis, through $K$, and find the relevant matrix elements of $M_*$ to get the intended P/NP formula. To our advantage, we don't have to determine every coefficient of $M_*$. We only need the ones that will be associated with the WKB-active actions, which are defined by the reduced actions of the periodic paths along the real-space axis, see, for example, Figure \textbf{\ref{fig:3}}.
\section{Classical and Quantum P/NP Relations}
\label{sec:PNP}
The Riemann bilinear relation and its encoding of the perturbative/nonperturbative relation are already known \cite{Basar:2017hpr, Codesido_2018}. However, there is no known explicit P/NP-relation formula for $\text{g}>1$ that relates the perturbative energy expansion $u(a,g)$ to the instanton function $A(a,g)$\cite{Gahramanov:2015yxk}. Now, we will lay out the general framework to derive a P/NP relation in general one-dimensional quantum mechanical systems with holomorphic potential energies $V(z)$ of genus-g that have degenerate minima when $V_{\text{min}}=0$ on the real line. The idea can be extended to systems with nondegenerate saddles straightforwardly.

Given a genus-g compact Riemann surface $X$ with the canonical basis $(\vec{\mathbf{a}},\vec{\mathbf{b}})$ and the corresponding canonical dissection $X_0$, any two Abelian differentials $\omega_1, \omega_2\in H^1(X,\mathbb{C})$ satisfy the \textit{Riemann bilinear identity} \cite{griffiths1989introduction,FarkasKra}
\begin{equation}
    \sum_{i=1}^{\text{g}}\left(\oint_{\mathbf{a}_i}\omega_1\oint_{\mathbf{b}_i}\omega_2-\oint_{\mathbf{b}_i}\omega_1\oint_\mathbf{a_i}\omega_2\right) = \int_X\omega_1\wedge\omega_2.
\end{equation}
Since $X_0$ is simply connected, there exists a function $f$ on $X_0$ such that $df=\omega_1$, then by Stokes' theorem one has
\begin{equation}
    \int_X\omega_1\wedge\omega_2 =\int_{X_0}d(f\omega_2)= \int_{\del X_0}f\omega_2=2\pi i\sum_{p\in\text{poles}}\text{Res}(f\omega_2,p).
\end{equation}
Using the symplectic intersection matrix the Riemann bilinear identity can be written in the compact form,
\begin{equation}
    \Pi_{\omega_1}^TJ\Pi_{\omega_2} = 2\pi i \sum_{z\in\text{poles}}\text{Res}(f\omega_2,z).
\end{equation}
In this form, it is apparent that the Riemann bilinear identity is invariant under $\text{Sp}(2\text{g},\mathbb{R})$. However, it is not always invariant under $\text{SL}(2\text{g},\mathbb{R})$ for $\text{g}\geq 2$ transformation. This fact, is in the core of the problem for finding an explicit quantum P/NP relation because the transformations $M_o\in \text{SL}(2\text{g},\mathbb{Z})$ that relates the PF-basis to WKB-active actions are almost always not an element of $\text{Sp}(2\text{g},\mathbb{Z})$ for $\text{g}\geq 2$. This fact gets even more convoluted together with the needed scaling transformations $S\in \text{GL}(2\text{g},\mathbb{R})$ and the fact that that PF-basis does not generically respect the canonical intersection structure itself. For non-canonical pairs, say PF-basis, the Riemann bilinear relation still holds as long as one uses the correct intersection structure
\begin{equation}
    \Pi_{\text{PF}}^TK\Pi_{\text{PF}} = 2\pi i\sum_{z\in\text{poles}}\text{Res}(f\omega_2,z).
\end{equation}

Therefore, to account for everything, we need to transform the intersection matrix $K$ into some other $K'$, not respecting the symplectic structure,
\begin{equation}\label{newinter}
    K = M_{*}^TK'M_{*}.
\end{equation}
This proves that the Riemann bilinear relation, by default, \textbf{does not} provide the physically relevant P/NP relation for the WKB-active cycles. However, we can still make use of the relation \eqref{newinter} to write a modified Riemann bilinear relation between the WKB-active actions in terms of the PF-basis. The modified Riemann bilinear relation, satisfied by the WKB-active actions $\Pi'_{\omega}$, gives
\begin{equation}
    \Pi_{\omega_1}'^TK'\Pi'_{\omega_2} = \Pi_{\omega_1}^TM_{*}^{T}K'M_{*}\Pi_{\omega_2} = \Pi_{\omega_1}^TK\Pi_{\omega_2}.
\end{equation}
Finally, it is now clear that our goal for finding a P/NP relation for a given system is to find the new intersection structure $K$ and then find the modified intersection matrix $K'$ via determining the relation between the PF-basis and WKB-active actions, i.e., finding a choice for $M_*$ that, in most cases, lives in the coset space $\text{GL}(2\text{g},\mathbb{R})/\text{Sp}(2\text{g},\mathbb{R})$ mapping the PF-basis actions to WKB-active actions. For systems exhibiting many symmetries, one may have $M_*\in \text{Sp}(2\text{g},\mathbb{Z})$ further simplifying the P/NP relation, e.g. the symmetric triple well, $V(q)= q^2(q-q_0)^2(q+q_0)^2$.  
\subsection{The Classical P/NP Relation}
The P/NP relation does not only stand for quantum mechanics. It also relates the classical actions of the complexified periodic orbits of a system. For the classical P/NP relation, we choose our differentials to be the classical action differentials  $\omega_1=p(q,u)dq$ and $\omega_2=\frac{\del}{\del u}p(q,u)dq$, where $p(q,u)=\pm\sqrt{2(u-V(q))}$ is the classical momentum, and choose the one-cycles to be some canonical basis $(\vec{\mathbf{a}},\vec{\mathbf{b}})$ so that we have the PF-basis actions and periods satisfying the Riemann bilinear relation, in a compact notational form,
\begin{equation}\label{classicalPNP1}
   \Pi_{pdq}^TJ\Pi_{dq/p}=\vec{a}\cdot\frac{\del \vec{a}_D}{\del u}-\vec{a}_D\cdot\frac{\del \vec{a}}{\del u} = \sum_{q_0\in\text{poles}}\text{Res}\left(f \frac{dq}{p},q_0\right).
\end{equation}
Notice that for $\text{g}=1$, this identity is nothing but the Wronskian for the Picard-Fuchs equation \eqref{PFequation} which is satisfied by the classical actions. Therefore, the Riemann bilinear relation is automatically satisfied by \textbf{any} solutions to the PF equation up to a constant factor. This statement is equivalent to the fact that $\text{SL}(2,\mathbb{Z}) = \text{Sp}(2,\mathbb{Z})$. 

For $\text{g} \geq 2$, the above is not always satisfied, and one would need the new intersection structure constants
\begin{equation}
    \Pi_{pdq}^T K\Pi_{dq/p}=\sum_{i=1}^{\text{g}}\kappa_i\left(a_i\frac{\del a_i^D}{\del u}-a_i^D\cdot\frac{\del a_i}{\del u}\right) = \sum_{z\in\text{poles}}\text{Res}\left(f \frac{dq}{p},z\right).
\end{equation}
Let us now use the fact that the relevant WKB-active actions $(\vec{W},\vec{W}_D)$ are related to the PF-basis actions with a linear transformation via \eqref{transform}. Using the matrix $M_o$, to get the modified Riemann bilinear relation we have the classical P/NP relation
\begin{equation}\label{classicalPNP}
    \Pi_{pdq}'^TK'\Pi'_{dq/p}=(\Pi_{pdq}^TM_*^{T})K'(M_*\Pi_{dq/p})=\sum_{q_0\in\text{poles}}\text{Res}\left(f \frac{dq}{p},q_0\right)
\end{equation}
This is the relation for the complexified classical mechanics, satisfied by the classical actions $\oint_{\mathbf{a}_i, \mathbf{b}_i} pdq$ in both classically allowed and forbidden regions. If one introduces the prepotential $\mathcal{F}(a)$ and use the definition
\begin{equation}
    \frac{\del \mathcal{F}(a)}{\del a} = a_D
\end{equation}
then the classical P/NP-relation is nothing but the so-called Matone relation \cite{MATONE1995342,Sonnenschein:1995hv,Eguchi:1995jh}.
\subsection{The Quantum P/NP Relation}
The idea in deriving the quantum P/NP relation is a straightforward application of the classical P/NP relation given in the previous subsection. In this case, we treat the coupling $g\neq 0$ as another moduli parameter, and choose the two quantum differentials to be the derivatives of the formally defined quantum action differentials with respect to the coupling $g$ and the energy parameter $u$,
\begin{equation}
 \omega_1 = \frac{\del}{\del g}p_Q(q,u,g)dq, \qquad    \omega_2=\frac{\del}{\del u}p_Q(q,u,g)dq
\end{equation}
respectively.
 Here, the anharmonic coupling constant $g$ is on equal footing with $\hbar$\footnote{One can switch to the perturbative normalization by scaling $u\to ug$.}. The Riemann bilinear identity for the PF-basis actions is
\begin{equation}
    \Pi^{T}_{\omega_1}K\Pi_{\omega_2}=\sum_{i=1}^{\text{g}}\kappa_i\left(\frac{\del a_i}{\del g}\frac{\del a_i^{D}}{\del u}-\frac{\del a_i^D}{\del g}\frac{\del a_i}{\del u}\right) = \sum_{q_0\in\text{poles}}\text{Res}\left(f\omega_2,q_0\right)
\end{equation}
where $K$ is the same intersection matrix found in the classical case.
The important note to make here is that some of the PF-basis dual actions, say $\vec{a}'_D$, corresponding to some $\vec{\textbf{b}}'$-cycles, will show up with logarithmic terms due to repeated roots of the indicial equation of the Picard-Fuchs equation \eqref{PFequation}. In general, the asymptotic structure of these PF basis actions $\vec{a}'_D(a)$ as a function of their duals will have the form
\begin{equation}\label{logform}
    a'_D(a,g) =A(a,g) + \log\Gamma\left(\frac{1}{2}+a\right) + a\log\left(\frac{g}{\mathcal{C}}\right) -\frac{1}{2}\log(2\pi)
\end{equation}
where $A(a,g)$ is the so-called \textit{instanton function} which is an asymptotic series in powers of $g$ whose coefficients are polynomials in $a$. Here, $\Gamma(x)$ is the Gamma function and the symbol $\mathcal{C}$ is a constant depending on the potential energy $V(q)$ of the system under consideration (for the anharmonic oscillators $V=\frac12q^2-q^d$ of Sec.~\ref{sec:examples}, with the analytic part of $a^{(0)}_{1,D}$ normalized as $-u+\dots$, one finds $\mathcal{C}=2^{d/(d-2)}$). The Gamma function is the familiar Weber factor of the local harmonic problem at the bottom of the well \cite{Dunne:2014bca,zinn-justin_multi-instantons_2004-1,Takei2}.

We will refer to the PF-basis action pairs of this type as the \textit{logarithmic pairs}. Depending on the symmetry structure of the saddles of $V(q)$, the number of logarithmic pairs will decrease. Therefore, not all $(a_i,a_i^D)$ are logarithmic pairs. In general, the number of logarithmic pairs in a system will amount to the number of topologically distinct nonperturbative actions in the system which are not related by monodromy transformations. This will increase the number of instanton fluctuation functions we need to describe the total spectrum of the system. Thus, we see that the inverted dual action's derivative with respect to $g$, treating $a$ and $g$ as independent variables, gives the simple expression\footnote{In practice, the inversion will happen only using the PF-basis action $a=u+\cdots$. For multi-well systems exhibiting multiple instanton configurations, this action will play the role of connecting all actions in the P/NP relations through $u(a)$.}
\begin{equation}\label{invA1}
    g\frac{\del a'_D(a,g)}{\del g} = g\frac{\del A(a,g)}{\del g}+a
\end{equation}

Most conveniently, the matrix $M_o$ found from the classical P/NP-relation \eqref{classicalPNP} is exactly the same transformation we use for the quantum P/NP-relation relating the WKB-active actions
\begin{equation}\label{quantumPNP}
    \Pi_{\omega_1}'^TK'\Pi_{\omega_2}' = \sum_{q_0\in\text{poles}}\text{Res}(f\omega_2,q_0)
\end{equation}
Together with the inverted relations of the instanton fluctuation functions \eqref{invA1}, the above equation describes the all-orders quantum P/NP-relation for arbitrary genus-g! Of course, the notation in which it is presented is only for visual appeal. Its intuitive understanding will become much clearer in the next section where we present some higher genus examples and explicitly do the algorithm of showing the P/NP relation between the instanton fluctuation functions and the perturbative actions.

Lastly, let us make some important remarks before presenting any examples. There are 2 important facts that have made an explicit formula for higher genus P/NP relations elusive until now:

The first one is the fact that in order to compute the modular invariant, the set of actions that go inside the Riemann bilinear relation has to have $2\text{g}$ linearly independent actions in it. In quantum mechanical systems that are easy to work with and have been studied extensively, the number of WKB-active actions is \textit{almost always} less than $2\text{g}$. Therefore, one cannot find a relationship between them without introducing the leftover actions, which complicates the process. On top of this difficulty, the Picard-Fuchs basis pairs also do not necessarily form canonical pairs. Therefore, one needs to introduce the topological data about their one-cycles and construct the correct intersection matrix.

Secondly, the Riemann bilinear relation is only invariant under $\text{Sp}(2\text{g},\mathbb{Z})$ and not necessarily under $\text{SL}(2\text{g},\mathbb{Z})$. The transformations from PF-basis to WKB-active actions almost never respect the intersection structure for $\text{g}\geq 2$; thus they are only part of $\text{SL}(2\text{g},\mathbb{Z})$ transformations. This implies that even if one has $2\text{g}$-many WKB-active actions, they won't satisfy the usual form of the Riemann bilinear relation, so one needs to consider the \textit{modified} bilinear relation \eqref{quantumPNP}.

\section{P/NP Relation for Higher Genus Anharmonic Oscillators}
\label{sec:examples}

In this section, we work out explicit higher-genus examples.  We will study the commonly known anharmonic oscillator potential \cite{Bender:1969si,Bender:1973rz,Alvarez2,jentschura_multi-instantons_2010}. Before we do so, we give a general remark on  their symmetries and  structure of critical points.  

\begin{figure}[t]
\centering
\begin{tikzpicture}[scale=0.92,>=stealth]
\definecolor{c0}{RGB}{31,119,180}\definecolor{c1}{RGB}{214,39,40}\definecolor{c2}{RGB}{44,160,44}
\definecolor{c3}{RGB}{255,127,14}\definecolor{c4}{RGB}{148,103,189}
\begin{scope}
\draw[gray!70,->] (-1.5,0)--(1.6,0) ;
\draw[gray!70,->] (0,-1.4)--(0,1.5) ;
\draw[gray!40,dashed] (0:1)--(72:1)--(144:1)--(216:1)--(288:1)--cycle;
\foreach \k/\c in {0/c0,1/c1,2/c2,3/c3,4/c4} {\fill[\c] (\k*72:1) circle (2.2pt); \node[\c] at (\k*72-17:1.32) {\scriptsize $q_{\k}$};}
\fill (0,0) circle (1.3pt);
\draw[->,gray,thick] (8:0.6) arc (8:64:0.6); \node[gray] at (36:0.42) {\scriptsize $R$};
\node[draw,gray,inner sep=1.5pt] at (-1.25,1.25) {\scriptsize $q$};
\node at (0,-1.85) {\small (a) septic, $m=5$: $q$-plane};
\end{scope}
\begin{scope}[xshift=3.9cm]
\draw[gray!70,->] (-1.5,0)--(1.6,0) ;
\draw[gray!70,->] (0,-1.4)--(0,1.5) ;
\draw[gray!40,dashed] (0,0) circle (1);
\foreach \k/\c in {0/c0,1/c1,2/c2,3/c3,4/c4} {\fill[\c] (\k*144:1) circle (2.2pt); \node[\c] at (\k*144-17:1.32) {\scriptsize $u_{\k}$};}
\fill (0,0) circle (1.3pt);
\node[draw,gray,inner sep=1.5pt] at (-1.25,1.25) {\scriptsize $u$};
\node at (0,-1.85) {\small $u$-plane, $\mathbb Z_5$};
\end{scope}
\begin{scope}[xshift=8.4cm]
\draw[gray!70,->] (-1.5,0)--(1.6,0) ;
\draw[gray!70,->] (0,-1.4)--(0,1.5) ;
\draw[gray!40,dashed] (0:1)--(60:1)--(120:1)--(180:1)--(240:1)--(300:1)--cycle;
\foreach \k/\c in {0/c0,1/c1,2/c2,3/c0,4/c1,5/c2} {\fill[\c] (\k*60:1) circle (2.2pt); \node[\c] at (\k*60-17:1.32) {\scriptsize $q_{\k}$};}
\fill (0,0) circle (1.3pt);
\draw[->,gray,thick] (8:0.6) arc (8:52:0.6); \node[gray] at (30:0.42) {\scriptsize $R$};
\node[draw,gray,inner sep=1.5pt] at (-1.25,1.25) {\scriptsize $q$};
\node at (0,-1.85) {\small (b) octic, $m=6$: $q$-plane};
\end{scope}
\begin{scope}[xshift=12.3cm]
\draw[gray!70,->] (-1.5,0)--(1.6,0) ;
\draw[gray!70,->] (0,-1.4)--(0,1.5) ;
\draw[gray!40,dashed] (0,0) circle (1);
\foreach \k/\c/\l in {0/c0/{0,3},1/c1/{1,4},2/c2/{2,5}} {\fill[\c] (\k*120:1) circle (2.2pt); \node[\c] at (\k*120-20:1.42) {\scriptsize $u_{\l}$};}
\fill (0,0) circle (1.3pt);
\node[draw,gray,inner sep=1.5pt] at (-1.25,1.25) {\scriptsize $u$};
\node at (0,-1.85) {\small $u$-plane, $\mathbb Z_3$};
\end{scope}
\end{tikzpicture}
\caption{Critical points $q_k$ of $V(q)$  and their critical values $u_k=u_*\zeta^{2k}$; the black dot is $q=0$, $u=0$. Colors follow each point to its image. 
(a) Septic: $u\mapsto\zeta^2u$ visits the five critical values in the order $0,3,1,4,2$ (counterclockwise), and only $q_0$ is real.
(b) Octic: the parity pairs $\pm q_k$ share a critical value, so $D_6$ acts on the $u$-plane as $D_3$, and $q_0$ and $q_3=-q_0$ are both real.}
\label{fig:sym}
\end{figure}
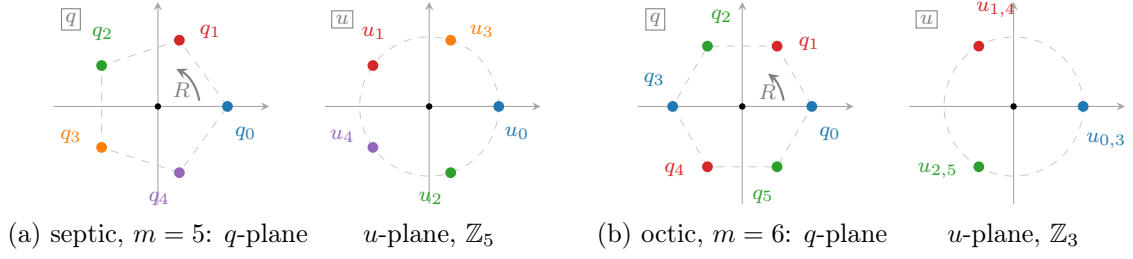

For $V(q)=\frac12q^2-q^d$ we set $m\equiv d-2$ and $\zeta\equiv e^{2\pi i/m}$. Since $q^d=q^{m+2}$, the curve
$p^2=2u-q^2+2q^d$ and the quantum Riccati equation are invariant under
\begin{equation}
R:\;(q,\,p,\,u,\,g)\;\longmapsto\;(\zeta q,\,\zeta p,\,\zeta^2u,\,\zeta^2g),\qquad p\,dq\;\longmapsto\;\zeta^2\,p\,dq ,
\label{eq:Zm}
\end{equation}
and, since $V$ is real, under complex conjugation $C$. With $CRC=R^{-1}$, these generate the dihedral group $D_m$ acting on the
$q$-plane. On $u$ and $p\,dq$ only $\zeta^2$ acts, so the effective group is $D_M$ with $M=m/\gcd(m,2)$. For even $d$, parity
$q\to-q$ is the central element $R^{m/2}$, which acts trivially on $u$.
The nontrivial critical points of $V$ are the vertices of a regular $m$-gon,
\begin{equation}
q_k=d^{-1/m}\zeta^k,\qquad u_k=V(q_k)=u_*\,\zeta^{2k}, 
\end{equation}
forming a single orbit of $D_m$ (Fig.~\ref{fig:sym}). Only $q_0$ is real for odd $d$, and only $\pm q_0$ for even $d$; these are the
barrier tops, and all other critical points are complex. Their critical values $u=0$ and $u^M=u_*^M$ are the singular points of the
Picard--Fuchs equations below. The instanton actions along the orbit differ only by phases, $S_k=\zeta^{2k}S_0$, with
$S_0=W^{(0)}_{\rm NP}(0)$.

\subsection{Sextic Oscillator}
Let us demonstrate the P/NP relation for the sextic anharmonic oscillator,
$$V(q) = \frac{1}{2}q^2-q^6.$$ 
We find the Picard-Fuchs relation for the classical actions to be
\begin{equation}\label{pfsextic}
    \left[u^2(-1+54u^2)\del_u^4 + 2u(-1+162u^2)\del_u^3+354u^2\del_u^2+30u\del_u\right]F^{(0)}(u) = 0.
\end{equation}
The linearly independent asymptotic series solutions are
\begin{align}\label{sexticPF}
    a_1^{(0)}(u) &= u+\frac{5 u^3}{2}+\frac{693 u^5}{16}+\frac{36465 u^7}{32}+\frac{37182145 u^9}{1024}+\frac{2646692685 u^{11}}{2048}+\mathcal{O}(u^{13})\\\nonumber
   a_2^{(0)}(u) &= u^2+\frac{32 u^4}{3}+\frac{3584 u^6}{15}+\frac{49152 u^8}{7}+\frac{74973184 u^{10}}{315}+\frac{872415232 u^{12}}{99}+\mathcal{O}(u^{14})\\\nonumber
   a_{1,D}^{(0)}(u)&=-u+\frac{17}{3}u^3  +\frac{18027}{160}u^5+\frac{1394891}{448}u^7 + \frac{3751204337}{36864}u^9 + a_1^{(0)}(u)\log\left(\frac{u}{2\sqrt{2}}\right)\\\nonumber
   a_{2,D}^{(0)}(u) &= 1
\end{align}
By applying the quantum generators to the general solutions, we find the higher order quantum corrections and generate the complete PF-basis actions $\Pi_{p_Qdq}=(\vec{a},\vec{a}_D)^t$ for the quantum action lattice $\Lambda_Q$, that is,
\begin{equation}
     \Pi_{p_Qdq} = \frac{1}{g}\sum_{n=0}^{\infty}g^n\mathcal{D}_n\Pi_{pdq}, \qquad \qquad   \Pi = \begin{pmatrix}
        \vec{a} \\
        \vec{a}_D \end{pmatrix}
 \end{equation}
Here, $g$ is on the same footing with $\hbar$. Recall that the PF-basis actions do not necessarily respect the symplectic structure, so they satisfy the bilinear identity with a new intersection matrix $K$,
\begin{equation}
    \Pi^T_{p_Qdq} K \Pi_{p_Qdq} = 0
\end{equation}
where the RHS is zero because the sum of all residues at infinity vanishes (see App.~\ref{app:residue}). Using the quantum PF-actions \eqref{sexticPF}, we find that
\begin{equation}
    K = \begin{pmatrix}
        0 & 0 &1&0\\
        0 & 0 &0 &\frac{1}{2}\\
        -1& 0 &0 &0\\
        0& -\frac{1}{2}&0 &0
    \end{pmatrix}.
\end{equation}
There are only two WKB-active actions, namely
\begin{equation}
    \WP(u,g) \coloneq a_1(u,g)
\end{equation}
and
\begin{equation}
    \WNP(u,g) \coloneq a_1^D(u,g) + \frac{\pi}{4\sqrt{2}}a_2^D(u,g).
\end{equation}
In order to determine the above linear combination coefficients, we integrate the classical action up to sufficiently large order along the nonperturbative WKB-active region. This allows us to fix the scaling of each action. For details on how to determine the linear combination for $\WNP$ and hence the matrix $M_*$, we refer to the appendices.

We still need the other two linearly independent PF-basis actions in our complete P/NP relation. We choose them to be fixed, then the WKB-active actions are related to the PF-basis actions through the following product of linear transformation $M_o\in \text{SL}(4,\mathbb{Z})$ and the scaling $S$, 
\begin{equation}
    S=\text{diag}\left(1,1,1,\frac{\pi}{4\sqrt{2}}\right),\quad M_o = \begin{pmatrix}
         1 & 0 &0 &0\\
         0 & 1 &0 &0\\
         0 & 0 &1 &1\\
         0 & 0 &0 &1
     \end{pmatrix} \implies M_oS = M_* = \begin{pmatrix}
         1 & 0 &0 &0\\
         0 & 1 &0 &0\\
         0 & 0 &1 &\frac{\pi}{4\sqrt{2}}\\
         0 & 0 &0 &\frac{\pi}{4\sqrt{2}}
     \end{pmatrix}.
 \end{equation}
 Then the modified symplectic intersection matrix is
 \begin{equation}
      K'= (M_*^T)^{-1}KM_*^{-1} = \begin{pmatrix}
         0& 0& 1& -1\\
        0& 0& 0& \frac{2\sqrt{2}}{\pi}\\
        -1&0& 0& 0\\
        1&-\frac{2\sqrt{2}}{\pi}& 0& 0
     \end{pmatrix}=\begin{pmatrix}
         0& 0& 1& -1\\
        0& 0& 0& \frac{4}{\pi}\cos\left(\frac{\pi}{4}\right)\\
        -1&0& 0& 0\\
        1&-\frac{4}{\pi}\cos\left(\frac{\pi}{4}\right)& 0& 0
     \end{pmatrix}
 \end{equation}
 Notice the terms of the modified intersection matrix depend purely on the trigonometric numbers of the $\mathbb{Z}_4$ symmetry.
 With it, we find the LHS of the modified Riemann bilinear relation 
\begin{align}
    \Pi'^T_{\omega_1}K'\Pi_{\omega_2}' = \frac{\del}{\del g}\begin{pmatrix}
        \WP\\
        a_2'\\
        \WNP\\
        a_2'^D
    \end{pmatrix}^T\begin{pmatrix}
        0& 0& 1& -1\\
        0& 0& 0& \frac{2\sqrt{2}}{\pi}\\
        -1&0& 0& 0\\
        1&-\frac{2\sqrt{2}}{\pi}& 0& 0
    \end{pmatrix}\frac{\del}{\del u}
    \begin{pmatrix}
        \WP\\
        a_2'\\
        \WNP\\
        a_2'^D
    \end{pmatrix}
\end{align}
\begin{equation}
     =-\frac{2\sqrt{2}}{\pi}\frac{\del a_2'^D}{\del g}\frac{\del a'_2}{\del u} + \frac{\del a_2'^D}{\del u}\frac{\del}{\del g}\left(\frac{2\sqrt{2}}{\pi}a'_2 - \WP\right) + \frac{\del \WP}{\del g}\frac{\del \WNP}{\del u} + \frac{\del \WP}{\del u}\frac{\del}{\del g}\left(a_2'^D - \WNP\right)\nonumber
\end{equation}
Substituting back $(a_2',a_2^{D}{'})$, we get the relation between the PF-basis and the WKB-active actions,
\begin{equation}
    g\frac{\del \WNP}{\del g}\frac{\del \WP}{\del u}+\frac{\pi}{4\sqrt{2}g}\frac{\del \WP}{\del u}-\frac{1}{2g}\frac{\del a_2}{\del u}=0
\end{equation}
After the inversion $u(\WP,g)$,
\begin{align}
    u\left(\WP,g\right)&=g \WP - g^3\left(\frac{25 \WP}{8} +\frac{5
   \WP^3}{2}\right)\\\nonumber&-g^5 \left(\frac{19277
   \WP}{256} +\frac{4145 \WP^3}{32}+\frac{393 \WP^5}{16}\right)+\mathcal{O}(g^7)
\end{align}
treating $\WP$ and $g$ as independent variables and using the logarithmic action's form \eqref{logform} we achieve the quantum P/NP-relation for the sextic oscillator in terms of the PF-basis,
\begin{align}
    g\frac{\del A}{\del g} +\WP + \frac{\pi}{4\sqrt{2}g}-\frac{1}{2g}\frac{\del a_2}{\del \WP} = 0
\end{align}
which gives us the desired expansion of the instanton function
\begin{align}
    A(\WP,g)&=\frac{\pi }{4 \sqrt{2} g}+g^2 \left(\frac{221
   W_P}{24}+\frac{17 W_P^3}{3}\right)\\&\nonumber
   +g^4 \left(\frac{2283899
   W_P}{7680}+\frac{38459 W_P^3}{96}+\frac{10727
   W_P^5}{160}\right)+\mathcal{O}(g^6)
\end{align}
After inverting $\WP(u,  g)$ back (and rescaling $u\to gu$ for the usual perturbative form) we get the perturbative form of the instanton function as a function of the energy parameter $u$ and the coupling $g$,
\begin{align}
    A(u,g) &= \frac{\pi }{4 \sqrt{2} g}+g^2 \left(\frac{221 u}{24}+\frac{17
   u^3}{3}\right)\\\nonumber&+g^4 \left(\frac{2504899 u}{7680}+\frac{45769
   u^3}{96}+\frac{17527 u^5}{160}\right) + \mathcal{O}(g^6)
\end{align}
The P/NP relation of the sextic oscillator is a remarkably explicit demonstration of quantum geometry aspect of resurgence. Even though the WKB-active actions do not involve the PF-basis action $a_2$ in their linear combination, their P/NP relation still involves this particular PF-basis action which, on the surface, has no contribution to the spectrum of the system. Thus, we see that each saddle of the system corresponding to a linearly independent reduced action, even at the classical level, contributes to the all-orders P/NP relation.
\subsection{Quintic Oscillator}

 Let us now demonstrate the P/NP relation for the quintic anharmonic potential
 \begin{equation}\label{quintic}
     V(q) = \frac{1}{2}q^2-q^5.
 \end{equation}
 The Picard-Fuchs equation for the classical action is
 \begin{align}\label{PF5}
     \Big[u(25000u^3-27)\del_u^4+(200000u^3-54)\del_u^3 + 317500u^2\del_u^2+35000u\del_u+\frac{6545}{2}\Big]F^{(0)}(u) = 0.
 \end{align}
 The general linearly independent asymptotic series solutions are 
 \begin{align} 
   a^{(0)}_1(u) &= u+\frac{315 u^4}{16}+\frac{692835
   u^7}{128}+\frac{9704539845
   u^{10}}{4096}+\mathcal{O}(u^{13})\\ \nonumber
   a^{(0)}_2(u) &= u^2+\frac{283309 u^5}{2592}+\frac{248945034845
   u^8}{6718464}+\frac{2534524784560098421
   u^{11}}{143667634176}+\mathcal{O}(u^{14})\\\nonumber
   a^{(0)}_{1,D}(u) &= -u+\frac{10865 u^4}{192}+\frac{78046343
   u^7}{4608}+\frac{2252653869793
   u^{10}}{294912}+\mathcal{O}(u^{13}) + a^{(0)}_1(u)\log\left(\frac{u}{2^{5/3}}\right)\\\nonumber
   a^{(0)}_{2,D}(u) &= 1+\frac{6545 u^3}{648}+\frac{1682469481
   u^6}{839808}+\frac{20544525471974735 u^9}{26121388032} + \mathcal{O}(u^{12})
 \end{align}
Applying the quantum generators to these solutions, we define the PF-basis actions as
 \begin{equation}
     \Pi_{p_Qdq} = \frac{1}{g}\sum_{n=0}^{\infty}g^n\mathcal{D}_n\Pi_{pdq}, \qquad \qquad   \Pi = \begin{pmatrix}
        \vec{a} \\
        \vec{a}_D \end{pmatrix}
 \end{equation}
 As in the case of sextic oscillator, the intersection matrix turns out to be
 \begin{equation}
     K = \begin{pmatrix}
         0&0&1&0\\
         0&0&0&\frac{1}{2}\\
         -1&0&0&0\\
         0&-\frac{1}{2}&0&0
     \end{pmatrix}.
 \end{equation}
 There is only one pair of WKB-active perturbative and nonperturbative actions that contribute to the spectrum which encircle the saddles on the real line. The linear combinations for the P and NP actions are found to be
 \begin{equation}
     W_{\text{P}}(u,g) \coloneq a_1(u,g)
 \end{equation}
 and
 \begin{equation}
     W_{\text{NP}}(u,g) \coloneq \frac{7}{(3\sqrt{3})^2}\frac{2^{1/3}\pi^2}{\Gamma^3(\frac{2}{3})}a_2(u,g)+a_1^D(u,g)
     +\frac{3\sqrt{3}}{7}\frac{\Gamma^3\left(\frac{2}{3}\right)}{2^{1/3}\pi}a_2^D(u,g).
 \end{equation}
 The above coefficients can be found by direct integration of the classical action (around NP cycle) up to $2\text{g}-1=3$ orders in $u$ (see appendices). By doing so, we fix the needed scaling coefficients of $S$,
 \begin{equation}
     S = \text{diag}\left(1, \frac{7}{(3\sqrt{3})^2}\frac{2^{1/3}\pi^2}{\Gamma^3(\frac{2}{3})},1,\frac{3\sqrt{3}}{7}\frac{\Gamma^3\left(\frac{2}{3}\right)}{2^{1/3}\pi}\right),
 \end{equation}
notice that $\det S \neq 1$, so its action on PF-basis already does not preserve their symplectic structure.

 We need 4 actions to be related by the modified Riemann bilinear relation so we leave the other 2 actions to be the scaled non-logarithmic PF-basis pair $(S_{22}a_2,S_{44}a_2^D)\equiv (a'_2,a'^D_2)$. Therefore, the linear transformation matrix $M_o\in \text{SL}(4,\mathbb{Z})$ connecting the PF-basis cycles to WKB-active cycles is
 \begin{equation}
     M_o = \begin{pmatrix}
         1 & 0 &0 &0\\
         0 & 1 &0 &0\\
         0 & 1 &1 &1\\
         0 & 0 &0 &1
     \end{pmatrix} \implies M_* = M_oS = \begin{pmatrix}
         1 & 0 &0 &0\\
         0 & \frac{7}{(3\sqrt{3})^2}\frac{2^{1/3}\pi^2}{\Gamma^3(\frac{2}{3})} &0 &0\\
         0 &  \frac{7}{(3\sqrt{3})^2}\frac{2^{1/3}\pi^2}{\Gamma^3(\frac{2}{3})} &1 &\frac{3\sqrt{3}}{7}\frac{\Gamma^3\left(\frac{2}{3}\right)}{2^{1/3}\pi}\\
         0 & 0 &0 &\frac{3\sqrt{3}}{7}\frac{\Gamma^3\left(\frac{2}{3}\right)}{2^{1/3}\pi}
     \end{pmatrix}
 \end{equation}
 The modified symplectic intersection matrix turns out to be
 \begin{equation}
     K' = (M_*^T)^{-1} K M_*^{-1} = \begin{pmatrix}
         0& -1& 1& -1\\
         1& 0& 0& \frac{3\sqrt{3}}{2\pi}\\
         -1& 0& 0& 0\\
         1& -\frac{3\sqrt{3}}{2\pi}& 0& 0
     \end{pmatrix}=\begin{pmatrix}
         0& -1& 1& -1\\
         1& 0& 0& \frac{3}{2\pi}\tan{\left(\frac{\pi}{3}\right)}\\
         -1& 0& 0& 0\\
         1& -\frac{3}{2\pi}\tan{\left(\frac{\pi}{3}\right)}& 0& 0
     \end{pmatrix}
 \end{equation}
 Again, notice that the elements of the matrix $K'$ only depend on the trigonometric values of its $\mathbb{Z}_3$ symmetry.
Then, the modified bilinear relation \eqref{quantumPNP} gives 
 \begin{equation}
      \Pi_{\omega_1}'^TK'\Pi_{\omega_2}' = \frac{\del}{\del g}\begin{pmatrix}
         \WP\\
         a'_2\\
         \WNP\\
         a_2'^D
     \end{pmatrix}^T\begin{pmatrix}
         0& -1& 1& -1\\
         1& 0& 0& \frac{3\sqrt{3}}{2\pi}\\
         -1& 0& 0& 0\\
         1& -\frac{3\sqrt{3}}{2\pi}& 0& 0
     \end{pmatrix}\frac{\del}{\del u}\begin{pmatrix}
         \WP\\
        a'_2\\
        \WNP\\
        a_2'^D
     \end{pmatrix}=0
     \end{equation}
     \begin{align}
     \implies&\frac{3\sqrt{3}}{2\pi}\frac{\del a_2'^D}{\del u}\left(\frac{\del a'_2}{\del g}-\frac{\del \WP}{\del g}\right) - \frac{3\sqrt{3}}{2\pi}\frac{\del a'_2}{\del u}\left(\frac{\del a_2'^D}{\del g}+ \frac{\del \WP}{\del g}\right) + \frac{\del \WP}{\del g}\frac{\del \WNP}{\del u}\\\nonumber&+ \frac{\del \WP}{\del u}\frac{\del}{\del g}\left(a_2' + a_2'^D - \WNP\right)  =0.\nonumber
 \end{align}
 Inverting $\WP(u,g)$ as $u(\WP,g)$,
  \begin{equation}
     u(\WP,g) = g \WP-g^4 \left(\frac{1107}{256}+\frac{1085
   }{32}\WP^2+\frac{315 }{16}\WP^4\right)+\mathcal{O}(g^7)
 \end{equation}
then treating $\WP$ and $g$ as independent variables, and using the logarithmic pair form \eqref{logform}, we get the quantum P/NP-relation between the perturbative series and the instanton function $A$,
\begin{equation}
    \frac{\del A}{\del g}+\frac{\WP}{g} - \frac{\del a'_2}{\del g}\left(\frac{3\sqrt{3}}{2\pi}\frac{\del a_2'^D}{\del \WP} + 1\right)  +\frac{\del a_2'^D}{\del g}\left(\frac{3\sqrt{3}}{2\pi}\frac{\del a'_2}{\del \WP}-1\right) = 0
\end{equation}
Rewriting the scaled actions in terms of the PF-basis actions, we get the nice result involving the linear combination coefficients explicitly,
 \begin{equation}
     g\frac{\del A}{\del g} +\WP+ g\frac{\del a_2^D}{\del g}\left(\frac{1}{2}\frac{\del a_2}{\del \WP}-\frac{7}{(3\sqrt{3})^2}\frac{2^{1/3}\pi^2}{\Gamma^3(\frac{2}{3})}\right) - g\frac{\del a_2}{\del g}\left(\frac{1}{2}\frac{\del a_2^D}{\del \WP}-\frac{3\sqrt{3}}{7}\frac{\Gamma^3(\frac{2}{3})}{2^{1/3}\,\pi}\right) = 0.
 \end{equation}
 We can find the instanton function $A(u,g)$ using the all-orders quantum P/NP relation, 
 \begin{align}
     A(\WP,g) &= \frac{3\sqrt{3}}{7}\frac{\Gamma^3\left(\frac{2}{3}\right)}{2^{1/3}\pi g}
     +\frac{7}{(3\sqrt{3})^2}\frac{2^{1/3}\pi^2}{\Gamma^3(\frac{2}{3})}\left(\frac{11}{28}+
     \WP^2\right)
   g\\\nonumber
   &+\frac{3\sqrt{3}}{7}\frac{\Gamma^3\left(\frac{2}{3}\right)}{2^{1/3}\pi }\left(\frac{2695 \WP}{288}+\frac{6545 \WP^3}{648}\right)
   g^2+\left(\frac{21171}{1024}+\frac{132245 \WP^2}{1152}+\frac{10865 \WP^4}{192}\right)
   g^3\\\nonumber &+\frac{7}{(3\sqrt{3})^2}\frac{2^{1/3}\pi^2}{\Gamma^3(\frac{2}{3})}
   \left(\frac{554677 \WP}{4608}+\frac{1218145 \WP^3
   }{5184}+\frac{181249 \WP^5}{2592}\right) g^4 + \mathcal{O}(g^5)
 \end{align}
 plugging back $\WP(u,g)$ and after the rescaling of the energy parameter $u\to gu$, we get the familiar perturbative result,
 \begin{align}
     A(u,g) &= \frac{3 \sqrt{3} \Gamma
   \left(\frac{2}{3}\right)^3}{7\cdot 2^{1/3} 
   \pi g}+\frac{7\cdot2^{1/3} \pi ^2
    }{\left(3
   \sqrt{3}\right)^2 \Gamma
   \left(\frac{2}{3}\right)^3}g\left(\frac{11}{28}+u^2\right)\\&+\frac{3
   \sqrt{3} \Gamma
   \left(\frac{2}{3}\right)^3
   }{7\cdot 2^{1/3} \pi }g^2\left(\frac{2695 u}{288}+\frac{6545
   u^3}{648}\right)+g^3
   \left(\frac{21171}{1024}+\frac{132245
   u^2}{1152}+\frac{10865
   u^4}{192}\right)\\&+\frac{7\cdot 2^{1/3}
   \pi ^2}{\left(3 \sqrt{3}\right)^2
   \Gamma \left(\frac{2}{3}\right)^3}g^4\left(\frac{594529
   u}{4608}+\frac{1569685
   u^3}{5184}+\frac{283309
   u^5}{2592}\right)+\mathcal{O}(g^{5}).
 \end{align}
 As demonstrated, the nonperturbative WKB-active action $\WNP$ is a linear combination of the three other PF-basis actions $\{a_1^D,a_2,a_2^D\}$ that are scaled by $S$ and the perturbative WKB-active action $\WP$ is exactly the other fourth PF-basis action $a_1$. Although $\WNP$ has all the other actions in it, the full size of the real vector space is 4-dimensional. Therefore, we cannot have a direct relation between $\WP$ and $\WNP$ without involving the other two linearly independent actions. Notice that such a choice is not unique; however, the choice made here involves the fewest extra steps to obtain a P/NP formula.
\subsection{Octic Oscillator}\label{sec:octic}
Let us present the P/NP relation for the first genus-3 potential. The potential is given by $V(q)=\frac{1}{2}q^2-q^8$. Similar to the sextic case, the system exhibits many symmetries (see Fig. \ref{fig:sym}). The PF equation for the classical action,
\begin{align}
    \Big[&124729\del^2_{u}+420096\del_u^3+299392u^2\del_u^4\\\nonumber&+4(-27+16384u^3)\del_u^5 + u(-27+4096u^3)\del_u^6\Big]\oint pdq = 0
\end{align}
Since the lowest order of the PF-operator is 2, we have a trivial solution $1$ and a linear solution $u$. The linearly independent solutions to this equation are the PF-basis actions
\begin{align}
    \Tilde{a}_1^{(0)}(u)&=u\\\nonumber
   \Tilde{a}_2^{(0)}(u)&=u^2+\frac{124729 u^5}{6480}+\frac{12738697499 u^8}{11757312}+\frac{693711249703043
   u^{11}}{8162933760}+\mathcal{O}(u^{14})\\\nonumber
   \Tilde{a}_3^{(0)}(u) &= u^3+\frac{21793 u^6}{648}+\frac{22007770015 u^9}{10077696}+\frac{13141865838321745
   u^{12}}{71833817088}+\mathcal{O}(u^{15})\\\nonumber
   \Tilde{a}_{1,D}^{(0)}(u) &= -u+\frac{2587 u^4}{288}+\frac{1242491 u^7}{2688}+\mathcal{O}(u^{10})+ \left(\Tilde{a}_1^{(0)}(u)+\frac{35}{8}\Tilde{a}_{3,D}^{(0)}(u)\right)\log\left(\frac{u}{2^{4/3}}\right)\\\nonumber
   \Tilde{a}_{2,D}^{(0)}(u)&= 1\\\nonumber
 \Tilde{a}_{3,D}^{(0)}(u)&= u^4+\frac{1287 u^7}{28}+\frac{1062347 u^{10}}{320}+\frac{300540195
   u^{13}}{1024}+\frac{954094903047 u^{16}}{32768}+\mathcal{O}(u^{19})\\\nonumber
\end{align}
It is clear that the combination $\Tilde{a}_1+\frac{35}{8}\Tilde{a}_{3}^D$ makes up the dual pair for $a_1^D$. With that, the combination $-\frac{455}{324}\Tilde{a}_3+\Tilde{a}_2^D$ becomes dual to $\Tilde{a}_2$. Therefore, we effectively have a genus-2 curve. We can redefine the PF-basis for this effective genus-2 system via
\begin{align}
    \begin{pmatrix}
        \vec{a}\\
        \vec{a}_D
    \end{pmatrix} = \begin{pmatrix}
             1&0&0&0&0&\frac{35}{8}\\
             0&0&-\frac{455}{324}&0&1&0\\
             0&0&1&0&0&0\\
             0&0&0&0&0&0\\
             0&1&0&0&0&0\\
             0&0&0&0&0&0
        \end{pmatrix}\begin{pmatrix}
            \vec{\Tilde{a}}\\
            \vec{\Tilde{a}}_D
        \end{pmatrix}
\end{align}
which gives
\begin{align}
    a_1^{(0)}(u) &= u+\frac{35 u^4}{8}+\frac{6435 u^7}{32}+O\left(u^9\right)\\\nonumber
    a_2^{(0)}(u) &= u^2+\frac{124729 u^5}{6480}+\frac{12738697499 u^8}{11757312}+\mathcal{O}(u^{11})
    \\\nonumber
    a_{1,D}^{(0)}(u) &= -u+\frac{2587 u^4}{288}+\frac{1242491 u^7}{2688}+\mathcal{O}(u^{10})+a_1^{(0)}(u)\log\left(\frac{u}{2^{4/3}}\right)\\
    a_{2,D}^{(0)}(u) &= 1-\frac{455 u^3}{324}-\frac{9915815 u^6}{209952}+O\left(u^9\right)
\end{align}
while keeping the pair $(a_3,a_3^D)$ the same.
Applying the quantum generators to these solutions, we define the PF-basis actions as
\begin{equation}
\Pi_{p_Qdq}=\frac1g\sum_{n=0}^{\infty}g^{n}\mathcal D_{n}\,\Pi_{pdq},\qquad \Pi=\begin{pmatrix}\vec a\\ \vec a_D\end{pmatrix}.
\end{equation}
Using the Riemann bilinear identity for the PF-basis, we find the intersection matrix to be
\begin{equation}
    K = \begin{pmatrix}
         0&0&0&1&0&0\\
    0&0&0&0&\frac{1}{2}&0\\
    0&0&0&0&0&0\\
    -1&0&0&0&0&0\\
    0&-1/2&0&0&0&0\\
    0&0&0&0&0&0
    \end{pmatrix}.
\end{equation}
Notice that the sector for the PF-basis pair $(a_3,a_3^D)$ is zero. Due to the symmetry of the system, one of the handles shrinks to zero, leading to an effective zero intersection between these cycles. Therefore, the problem of finding P/NP simplifies drastically, effectively, to a genus-2 case.
WKB-active actions $(\WP,\WNP)$ have the following linear combinations,
\begin{align}
    \WP(u,g) &= a_1(u,g)\nonumber\\
    \WNP(u,g) &= a_1^D(u,g) -\frac{\sqrt{3}\Gamma^3(1/3)}{2^{5/3}\cdot5\pi}a_2^D(u,g) -\frac{5\pi^2}{2^{1/3}\cdot9\Gamma^3(1/3)}a_2(u,g)
\end{align}
Thus, the scaling matrix and the corresponding modular transformation have to be
\begin{equation}
    S = \text{diag}\left(1,\frac{5\pi^2}{2^{1/3}\cdot9\Gamma^3(1/3)},1,\frac{\sqrt{3}\Gamma^3(1/3)}{2^{5/3}\cdot5\pi}\right),\quad M_o = \begin{pmatrix}
        1&0&0&0&0\\
        0&1&0&0&0\\
        0&-1&0&1&-1\\
        0&0&0&0&1
    \end{pmatrix}
\end{equation}
Using the product matrix $M_*=M_oS$, we get the modified intersection matrix $K' = M_*^TKM_*$, 
\begin{equation}
    K' = \begin{pmatrix}
        0&1&1&1\\
        -1&0&0&\frac{6}{\pi}\cos(\pi/6)\\
        -1&0&0&0\\
        -1&-\frac{6}{\pi}\cos(\pi/6)&0&0
    \end{pmatrix}.
\end{equation}
We recognize the cancellation of all gamma terms and the coefficients that purely depend on the trigonometric identities of the $\mathbb{Z}_6$ symmetry of its critical points (Fig. \ref{fig:sym}).

Finally, the modified bilinear relation gives the desired P/NP relation
\begin{align}
    g\frac{\del A}{\del g}+\WP - g\frac{\del a_2}{\del g}\left(\frac{1}{2}\frac{\del a_2^D}{\del \WP} - \frac{5\pi^2}{9\cdot2^{1/3}\Gamma^3(1/3)}\right) + g\frac{\del a_2^D}{\del g}\left(\frac{1}{2}\frac{\del a_2}{\del \WP}+\frac{\sqrt{3}\Gamma^3(1/3)}{5\cdot 2^{5/3}\pi}\right) =0
\end{align}
which reads
\begin{align}
    A(W_P,g)={}&\frac{\sqrt{3}\Gamma^3(1/3)}{5\cdot 2^{5/3}\pi g}+ \frac{5\pi^2}{9\cdot2^{1/3}\Gamma^3(1/3)}\,g\Big(\frac{17}{20}+W_P^2\Big)\nonumber\\&-\frac{\sqrt{3}\Gamma^3(1/3)}{5\cdot 2^{5/3}\pi}\,g^2\Big(\frac{385}{144}W_P+\frac{455}{324}W_P^3\Big)\nonumber\\
&+g^3\Big(\frac{28007}{2560}+\frac{22669}{576}W_P^2+\frac{2587}{288}W_P^4\Big)+\mathcal{O}(g^4),
\end{align}
after the inversion and the scaling for the perturbative form, we get
\begin{align}
    A(u,g)={}&\frac{\sqrt{3}\Gamma^3(1/3)}{5\cdot 2^{5/3}\pi g}+\frac{5\pi^2}{9\cdot2^{1/3}\Gamma^3(1/3)}\,g\Big(\frac{17}{20}+u^2\Big)-\frac{\sqrt{3}\Gamma^3(1/3)}{5\cdot 2^{5/3}\pi}\,g^2\Big(\frac{385}{144}u+\frac{455}{324}u^3\Big)\nonumber\\
&+g^3\Big(\frac{28007}{2560}+\frac{22669}{576}u^2+\frac{2587}{288}u^4\Big)\nonumber\\
&+\frac{5\pi^2}{9\cdot2^{1/3}\Gamma^3(1/3)}\,g^4\Big(\frac{839069}{11520}u+\frac{279565}{2592}u^3+\frac{124729}{6480}u^5\Big).
\end{align}
\subsection{Septic Oscillator}
Lastly, let us demonstrate the P/NP relation for the septic oscillator. The potential is given by $V(q)=\frac{1}{2}q^2-q^7$. The PF equation for the classical action is
\begin{align}
    &\Big[2u\,(26353376\,u^5-3125)\,\del_u^6+8\,(158120256\,u^5-3125)\,\del_u^5+9571922640\,u^4\del_u^4\\\nonumber&+26315728320\,u^3\del_u^3\,
+22054750452\,u^2\del_u^2+2156405328\,u\,\del_u-44920953\Big]F^{(0)}(u)=0,
\end{align}
Solving the Picard-Fuchs equation, we get the PF-basis
\begin{align}
    a_1^{(0)}(u)&=u +\frac{3003 u^{6}}{32} +\frac{456326325 u^{11}}{2048} +\frac{456375395290815 u^{16}}{524288}+\mathcal O(u^{21})\nonumber\\
a_2^{(0)}(u)&=u^{2} +\frac{767895089 u^{7}}{1500000} +\frac{46439331453649868279 u^{12}}{33000000000000}+\mathcal O(u^{17})\nonumber\\
a_3^{(0)}(u)&=u^{3} +\frac{14126055603 u^{8}}{14000000} +\frac{6226381217641869229 u^{13}}{2000000000000}+\mathcal O(u^{18})\nonumber\\
a_{1,D}^{(0)}(u)&=-u +\frac{2184119 u^{6}}{9600} +\frac{520166208311 u^{11}}{901120}+\mathcal O(u^{16})+a_1^{(0)}(u)\log\Big(\frac{u}{2^{7/5}}\Big)\nonumber\\
a_{2,D}^{(0)}(u)&=1 -\frac{14973651 u^{5}}{1000000} -\frac{29324077487611137 u^{10}}{1000000000000}+\mathcal O(u^{15})\nonumber\\
a_{3,D}^{(0)}(u)&=u^{4} +\frac{18008592521 u^{9}}{12000000} +\frac{1454097237841629766677 u^{14}}{286000000000000}+\mathcal O(u^{19})
\end{align}
Applying the quantum generators to these solutions, we define the PF-basis actions as
\begin{equation}
\Pi_{p_Qdq}=\frac1g\sum_{n=0}^{\infty}g^{n}\mathcal D_{n}\,\Pi_{pdq},\qquad \Pi=\begin{pmatrix}\vec a\\ \vec a_D\end{pmatrix}.
\end{equation}
Using the Riemann bilinear identity for the PF-basis, we find the intersection matrix to be
\begin{equation}
    K=\begin{pmatrix}
    0&0&0&1&0&0\\
    0&0&0&0&\frac{1}{2}&0\\
    0&0&0&0&0&\frac{3949491}{200000}\\
    -1&0&0&0&0&0\\
    0&-1/2&0&0&0&0\\
    0&0&-\frac{3949491}{200000}&0&0&0
    \end{pmatrix}\nonumber\\
\end{equation}
Again, there are only two WKB-active actions, 
\begin{equation}
    \WP(u,g)\coloneq a_1(u,g)
\end{equation}
\begin{align}
    \WNP = a_1^D(u,g)+c_2\,a_2(u,g)+c_3\,a_3(u,g)+c_2^D\,a_2^D(u,g)+c_3^D\,a_3^D(u,g),
\end{align}
where the coefficients are 
\begin{align}
    c_2&=\frac{9}{10}\,2^{-2/5}\,\frac{\Gamma^2(\frac35)}{\Gamma(\frac15)},&
    c_2^D=\frac{2}{9}\,2^{-3/5}\,\frac{\Gamma^2(\frac25)}{\Gamma(\frac45)},&
\nonumber\\
c_3&=-\frac{299}{1000}\,2^{1/5}\,\frac{\Gamma^2(\frac15)}{\Gamma(\frac25)},&\nonumber
c_3^D=\frac{13209}{2000}\,2^{4/5}\,\frac{\Gamma^2(\frac45)}{\Gamma(\frac35)}.
\end{align}
Therefore, the scaling matrix and the corresponding modular transformation have to be
\begin{equation}
    S=\text{diag}\left(1, c_2,c_3,1,c_2^D,c_3^D\right)
\end{equation}
\begin{equation}
M_o=\begin{pmatrix}1&0&0&0&0&0\\0&1&0&0&0&0\\0&0&1&0&0&0\\0&1&1&1&1&1\\0&0&0&0&1&0\\0&0&0&0&0&1\end{pmatrix}\in SL(6,\mathbb Z)\nonumber\\
\implies M_*=M_oS=\begin{pmatrix}1&0&0&0&0&0\\0&c_2&0&0&0&0\\0&0&c_3&0&0&0\\0&c_2&c_3&1&c_2^D&c_3^D\\0&0&0&0&c_2^D&0\\0&0&0&0&0&c_3^D\end{pmatrix}.
\end{equation}
The modified symplectic intersection matrix turns out to be 
\begin{align}
    K' =& \begin{pmatrix}0&-1&-1&1&-1&-1\\1&0&0&0&\frac{1}{2c_2c_2^D}&0\\1&0&0&0&0&\frac{3949491}{200000}\frac{1}{c_3c_3^D}\\-1&0&0&0&0&0\\1&-\frac{1}{2c_2c_2^D}&0&0&0&0\\1&0&-\frac{3949491}{200000}\frac{1}{c_3c_3^D}&0&0&0\end{pmatrix}\\&\nonumber= \begin{pmatrix}0&-1&-1&1&-1&-1\\1&0&0&0&\frac{5}{2\pi}\tan(2\pi/5)&0\\1&0&0&0&0&\frac{5}{2\pi}\tan\left(4\pi/5\right)\\-1&0&0&0&0&0\\1&-\frac{5}{2\pi}\tan(2\pi/5)&0&0&0&0\\1&0&-\frac{5}{2\pi}\tan(4\pi/5)&0&0&0\end{pmatrix}
\end{align}
which gives the desired P/NP relation
\begin{align}
&g\frac{\del A}{\del g} + \WP -g \frac{\del a_2}{\del g}\left(\frac{1}{2}\frac{\del a_2^D}{\del \WP} + c_2\right)+g \frac{\del a_2^D}{\del g}\left(\frac{1}{2}\frac{\del a_2}{\del \WP} - c_2^D\right)\\\nonumber & - g\frac{\del a_3}{\del g}\left(\frac{3949491}{200000}\frac{\del a_3^D}{\del \WP} +c_3\right) +g \frac{\del a_3^D}{\del g}\left(\frac{3949491}{200000}\frac{\del a_3}{\del \WP} - c_3^D\right) = 0
\end{align}
From which, one can derive the instanton function 
\begin{align}
    A(\WP,g)=&\frac{2^{2/5}\Gamma^2(\frac25)}{9\Gamma(\frac45)g}+\frac{9}{2^{3/5}\cdot5}\frac{\Gamma^2(\frac35)}{\Gamma(\frac15)}\left(\WP^2+\frac{25}{36}\right)g\\&-\frac{299}{1000}\,2^{1/5}\,\frac{\Gamma^2(\frac15)}{\Gamma(\frac25)}\left(\WP^3+\frac{145\,\WP}{92}\right)g^2\nonumber\\&+\frac{13209}{2000}\,2^{4/5}\,\frac{\Gamma^2(\frac45)}{\Gamma(\frac35)}\left(\WP^4+\frac{215\,\WP^2}{74}+\frac{2485}{5328}\right)g^3\nonumber\\
&-\frac{2^{2/5}\Gamma^2(\frac25)}{9\Gamma(\frac45)}\Big(\frac{5463843 \WP}{128000} +\frac{5578419 \WP^{3}}{80000} +\frac{14973651 \WP^{5}}{1000000}\Big)g^4\nonumber\\
&+\Big(\frac{32487253}{73728} +\frac{453063233 \WP^{2}}{153600} +\frac{13613173 \WP^{4}}{7680} +\frac{2184119 \WP^{6}}{9600}\Big)g^5+\mathcal O(g^6),\nonumber
\end{align}
and after reinverting $u(\WP,g)$ and scaling $u\to ug$, we get the perturbative result for the instanton function
\begin{align}
    A(u,g) &= \frac{2^{2/5}\Gamma^2(\frac25)}{9\Gamma(\frac45)g}+\frac{9}{2^{3/5}\cdot5}\frac{\Gamma^2(\frac35)}{\Gamma(\frac15)}\left(u^2+\frac{25}{36}\right)g\\&\nonumber-\frac{299}{1000}\,2^{1/5}\,\frac{\Gamma^2(\frac15)}{\Gamma(\frac25)}\left(u^3+\frac{145\,u}{92}\right)g^2\\\nonumber&+\frac{13209}{2000}\,2^{4/5}\,\frac{\Gamma^2(\frac45)}{\Gamma(\frac35)}\left(u^4+\frac{215\,u^2}{74}+\frac{2485}{5328}\right)g^3\nonumber\\
&-\frac{2^{2/5}\Gamma^2(\frac25)}{9\Gamma(\frac45)}\Big(\frac{5463843 u}{128000} +\frac{5578419 u^{3}}{80000} +\frac{14973651 u^{5}}{1000000}\Big)g^4\nonumber\\
&+\Big(\frac{32487253}{73728} +\frac{453063233 u^{2}}{153600} +\frac{13613173 u^{4}}{7680} +\frac{2184119 u^{6}}{9600}\Big)g^5+\mathcal{O}(g^6). \nonumber
\end{align}
\section{From anharmonic oscillators to multi-well potentials}\label{sec:multiwell}
The anharmonic oscillators of Sec.~\ref{sec:examples} are, in a sense, the most economical higher-genus systems. They have a single instanton type and a single WKB-active pair $(\WP,\WNP)$, and the remaining PF-basis actions enter the P/NP relation only as passive partners, through $K'$ and through the linear combination $\WNP$. Multi-well potentials
\begin{equation}
    V(q)=\tfrac{1}{2}\mathcal{W}'(q)^2,\qquad \mathcal{W}'(q)=\prod_{j=1}^N(q-q_j),\qquad q_1<q_2<\dots<q_N,
    \label{eq:multiwell}
\end{equation}
with $N$ degenerate minima and $N-1$ topologically distinct instantons \cite{Dunne:2020gtk}, sit at the opposite end. Their curve $p^2=2u-\mathcal{W}'(q)^2$ has genus $N-1$, and, unlike for the anharmonic oscillators, \textit{every cycle is physical}. In a companion paper \cite{Ture:2026mw} we show the following.
\begin{itemize}
\item {\bf The pairing is fixed by topology.} The barrier $\beta_k$ between the wells $k$ and $k+1$ pairs with the alternating sum $\alpha_k=\gamma_k-\gamma_{k-1}+\gamma_{k-2}-\dots$ of the wells $\gamma_j$ on its left. Together they form a symplectic basis, so that $M_*\in\text{Sp}(2\text{g},\mathbb{Z})$ and the modified intersection matrix is the canonical one, $K'=J$: no passive actions appear.
\item {\bf A quantum Riemann bilinear identity with a classical right-hand side.} With $a_k$ and $a_{k}^D$ the quantum actions of $\alpha_k$ and $\beta_k$, and $\{f,h\}=\del_gf\,\del_uh-\del_gh\,\del_uf$,
\begin{equation}
    \sum_{k=1}^{N-1}\{a_k,a_{k}^D\}=\frac{\mathcal{S}}{g^3},\qquad
    \mathcal{S}=2\,\text{Res}_{q=\infty}\frac{\mathcal{W}\,dq}{\mathcal{W}'}=\sum_{k=1}^{N-1}\frac{S_k}{\Omega_k}=\frac{2\sum_{i<j}(q_i-q_j)^2}{(N-1)N(N+1)},
    \label{eq:mwmain}
\end{equation}
where $S_k$ is the instanton action of barrier $k$ and an effective frequency $\Omega_k$ is defined by $\frac{1}{\Omega_k}=|\sum_{j\leq k}\frac{(-1)^j}{\omega_j}|
$. For $N=2$, $\mathcal{S}=\frac{1}{3}$ and \eqref{eq:mwmain} is the P/NP relation of the double well \cite{Dunne:2013ada,Dunne:2014bca}. 
\item {\bf $\Gamma$-structure and spectra.} Each barrier action carries one Weber factor $\Gamma(\frac12+B_j)$ for each of the two wells it connects, and all Gamma functions drop out of \eqref{eq:mwmain}. For three wells the relation becomes a two-sector P/NP relation in which the classical constant is shared between the sectors.
\end{itemize}
The two classes of examples are complementary. For the anharmonic oscillators the right-hand side of \eqref{quantumPNP} vanishes identically, as a consequence of their scaling symmetry (App.~\ref{app:residue}), and the P/NP content is carried by the passive PF-basis actions and the trigonometric entries of $K'$. For multi-well potentials there are no passive actions, $K'=J$, and the P/NP content is carried by the residue at infinity itself, exactly as in genus one. In both cases the right-hand side is purely classical, and the modified Riemann bilinear identity of Sec.~\ref{sec:PNP} is the common language.

\section{Conclusion}
\label{sec:conclusion}

In this work, we developed a general framework for constructing the
P/NP relation in one-dimensional
quantum mechanical systems with potentials of arbitrary genus 
using tools from complex algebraic geometry combined with exact WKB analysis.

One of the
central obstructions to a naive genus-g generalization of the genus-1
story is that the physically relevant WKB-active actions are related to
the natural Picard-Fuchs basis by a linear transformation in
$SL(2\text{g},\mathbb{Z})$ that generically fails to preserve the symplectic
intersection pairing once $\text{g}\geq2$. This is unlike in the genus-1 case, where
$\text{SL}(2,\mathbb{Z})$ and $\text{Sp}(2,\mathbb{Z})$ exactly coincide.  
We demonstrated that this
obstruction can be systematically absorbed into a modified intersection
matrix $K'$, yielding a modified Riemann bilinear identity satisfied by
the WKB-active actions rather than by the Picard-Fuchs basis only.
Promoting this identity to the quantum level, with the coupling $g$ and
energy $u$ treated as independent moduli, produces an all-orders P/NP
relation valid for arbitrary genus.

We demonstrated the method explicitly on four anharmonic oscillators, 
the genus-2 quintic and sextic potentials, and the genus-3 septic and octic
potentials,  deriving both the governing Picard-Fuchs equations and
their full solution bases. As a check, the resulting
instanton fluctuation functions $A(u,g)$ agree with results available
independently in the literature (e.g.\ the 
computations of Zinn-Justin and Jentschura \cite{zinn-justin_multi-instantons_2004-1,zinn-justin_multi-instantons_2004, jentschura_multi-instantons_2010}, obtained without
reference to any P/NP structure). We emphasize,
however, that the genuine novelty of this work does not lie in these
individual coefficients -- which were already known -- but in
establishing, for the first time, the existence and explicit
closed-form of a P/NP relation for genus $\text{g}>1$ potentials.

A notable
structural feature
that emerged in these examples is the direct link between the
multiplicity of the Picard-Fuchs indicial roots and the logarithmic
pairs: whenever the indicial equation has a
repeated root, the corresponding Frobenius solution necessarily
acquires a logarithm, and it is exactly this pair of actions -- one
regular, one logarithmic -- that carries the non-perturbative,
instanton-fluctuation content of the spectrum. Non-repeated roots, by
contrast, give rise to non-logarithmic action pairs with no associated
monodromy at the origin of moduli space. This gives a clean
algebraic-geometric criterion, visible already at the level of the
classical Picard-Fuchs equation, for identifying which cycles in a
higher-genus system are physically tied to instanton sectors before
any WKB or quantum analysis is performed.

Several directions remain open. 

{\bf Multi-well systems with topologically distinct instantons:} Our examples were restricted to
potentials with a single instanton type; extending the construction to
genuinely multi-well potentials with several topologically distinct
instantons, where multiple logarithmic pairs and coupled Picard-Fuchs
sectors appear simultaneously, is a natural next step. We take this step in a companion paper \cite{Ture:2026mw}, whose main result is summarized in Sec.~\ref{sec:multiwell}.  This is closely tied to a second question: understanding when the
transformation matrix $M_o$ can be chosen (or is forced) to lie in
$\text{Sp}(2g,\mathbb{Z})$, simplifying the P/NP relation considerably, as in
the symmetric triple well. Indeed the symmetry of $V(q)$ responsible
for $M_o$ landing in $\text{Sp}(2\text{g},\mathbb{Z})$ is precisely what controls how
many independent logarithmic pairs -- and hence how many distinct
instanton sectors -- a given multi-well potential can support: the
multi-well generalization and the $M_o\in \text{Sp}(2g,\mathbb{Z})$ question
are, in this sense, two facets of the same underlying problem, tied
together by the indicial-root multiplicity structure identified here.

{\bf Connection to supersymmetric gauge theory:}  Since the genus-1 P/NP
relation has a well-understood connection to Seiberg-Witten theory and
the Nekrasov-Shatashvili limit \cite{Seiberg:1994rs,Nekrasov:2009rc, Mironov:2009uv,Gaiotto:2009hg,Ito:2017iba}, it is natural to ask whether the
higher-genus modified bilinear structure identified here has an
analogous gauge-theoretic or topological-string interpretation,
potentially connecting to genus-g Seiberg-Witten curves and their
associated special geometry.  We leave these questions for future work.

\vspace{20pt}
\noindent\textbf{Acknowledgments.}
M. \"U. is supported by U.S. Department of Energy, Office of Science, Office of Nuclear Physics
under Award Number DE-FG02-03ER41260 and by the Simons Foundation Grant (Simons
Collaboration on Confinement and QCD Strings).
 

\appendix

\section{Derivation of the Picard-Fuchs equation and the quantum generators}\label{app:PF}
In this section we give a straightforward method of deriving the Picard-Fuchs equation for the classical action as well as the coefficients of the quantum generators. The method presented here, although less elegant, is equivalent to the Griffiths-Dwork algorithm \cite{Griffiths:1969a,Griffiths:1969b,griffiths1989introduction}.

Let $X$ be the compact Riemann surface defined by the (hyper)elliptic curve
\begin{equation}
    p^2 = 2(u-V(q))
\end{equation}
of genus g where $\text{g} = \lfloor\frac{\text{deg}\,V(q)-1}{2}\rfloor$. The first deRham cohomology space of $X$ is spanned by $2\text{g}$-many Abelian differentials. Therefore, the classical action differential $\omega_0=pdq\in H^{1}_{\text{dR}}(X,\mathbb{C})$ can be expressed as a linear combination of $ 2\text{g}$ differential one-forms, up to an exact form. One can generate these one-forms by taking partial derivatives of $\omega_0$ with respect to the moduli parameter u, i.e., 
\begin{equation}
    \sum_{k=0}^{2\text{g}}f_k(u)\frac{\del^k}{\del u^k}\omega_0 + dF = 0
\end{equation}
Integrating over any nonvanishing cycle $\gamma \in H_1(X,\mathbb{Z})$ one obtains the Picard-Fuchs equation for the classical actions
\begin{equation}
    \sum_{k=0}^{2\text{g}} f_k(u)\frac{\del^k}{\del u^k}\oint_{\gamma}\omega_0 = 0
\end{equation}
    In order to find the coefficients $f_k(u)$ that make up the Picard-Fuchs operator $\mathcal{L}_0$, we express the exact form, $dF$, as a linear combination of total derivatives of closed forms
\begin{equation}
    \sum_{k=0}^{2\text{g}}f_k(u)\frac{\del^k}{\del u^k}p - \sum_{n=1}^{n_{\text{max}}}g_n(u)\frac{d}{dq}\left(\frac{q^{n}}{p^{4g-3}}\right) = 0
\end{equation}
where $f_k(u)$ and $g_n(u)$ are polynomials in $u$ and $n_{\text{max}}$ is sufficiently large so that there is a sufficient number of equations to be solved. 
After multiplying with a common denominator one gets a system of linear equations for $f_k(u)$ and $g_n(u)$'s by setting each order in powers of q to zero. The largest order $f_{2\text{g}}(u)$ is always equal to the discriminant of $\Delta_{q}(\Sigma)$ of the (hyper)elliptic curve $\Sigma$, capturing the singular points of the moduli space in its roots. Then one gets the Picard-Fuchs differential operator for the classical action
\begin{equation}
    \mathcal{L}_0 \equiv \sum_{k=0}^{2\text{g}}f_{k}(u)\frac{\del^k}{\del u^k
    }
\end{equation}

\noindent
{\bf Quantum generators:} A similar process can be implemented for finding the coefficients of the quantum generators, $\mathcal{D}_n$, generating the $n^{\text{th}}$ quantum correction $\omega_{n} = \frac{\rho_n}{p^{3n-1}}dq$. For that, one needs to solve the system of linear equations
\begin{equation}
    \sum_{k \in P_n[\text{g}]}h_k^{(n)}(u)\frac{\del ^k}{\del u^k}p - \frac{\rho_n}{p^{3n-1}} - \sum_{m=1}^{m_{\text{max}}}r_{m}^{(n)}(u)\frac{d}{dq}\left(\frac{q^{m}}{p^{2gn-5}} \right) = 0 
\end{equation}
for $h_k^{(n)}(u)$ and $r_m^{(n)}(u)$'s where the index set $P_n[\text{g}] = \left\{(n-2)\text{g}, \cdots, n\text{g} -1\right\}$ is defined the same as in the main text. The indexing $P_n[\text{g}]$ is chosen so that the coefficients $h_k^{(n)}(u)$ stay as polynomials in $u$ rather than rational functions of $u$. Thus, one finds the quantum generators 
\begin{equation}
    \mathcal{D}_{n} \equiv \sum_{k \in P_n[\text{g}]}h_k^{(n)}(u)\frac{\del ^k}{\del u^k}.
\end{equation}
We should remark that the quantum generators are not unique. One can generate all types of different differential operators for a given one-form.
\subsection{Quantum generators of anharmonic oscillators and the quantum corrections}
Here we present the lists of quantum generator coefficients for each of the oscillator examples given in the manuscript as the pages permit.
\subsubsection{Sextic Oscillator}
For the quantum generators of the sextic oscillator, one can use the indexing $P_n[\text{g}]$ defined as above to determine the coefficients. An attempt of doing so will show that one would always get one out of the four coefficients to be zero. This has to do with the number of symmetries the sextic oscillator's saddles exhibit. A similar finding was also apparent in the number of nonconstant PF-basis actions it had. Therefore, to utilize this as a computational advantage, we look for 3 coefficients for each quantum generator operator $\mathcal{D}_{2n}$ as
\begin{equation}
    \mathcal{D}_{n} \coloneq \sum_{k=3n/2-2}^{3n/2}h_k^{(n)}(u)\frac{\del^k}{\del u^k}
\end{equation}
where for each $n$, we find the operators to be (up to $n=6$),
\begin{flalign}
    &\mathcal{D}_2 =-\frac{45 u}{8}\del_u+\left(\frac{13}{48}-\frac{315 u^2}{8}\right)\del^2_u+\left(\frac{5
   u}{16}-\frac{135 u^3}{8}\right)\del^3_u&
\end{flalign}
\begin{flalign}
    &\mathcal{D}_4= \left(\frac{65 u^2}{1408}+\frac{1327}{92160}\right)\del^4_u+\left(\frac{1755
   u^3}{11264}+\frac{133 u}{67584}\right)\del^5_u+\left(\frac{369 u^4}{11264}-\frac{41
   u^2}{67584}\right)\del_u^6&\nonumber
\end{flalign}
\begin{flalign}
    \mathcal{D}_6 &= \left(\frac{200875 u^3}{6684672}-\frac{18345121
   u}{72194457600}\right)\del_u^7+\left(\frac{4255011 u^4}{545914880}-\frac{48356671
   u^2}{393058713600}+\frac{91003}{178362777600}\right)\del_u^8\nonumber\\&+\left(\frac{218919
   u^5}{545914880}-\frac{496219 u^3}{43673190400}+\frac{91003
   u}{1248539443200}\right)\del_u^9\nonumber&
\end{flalign}
The perturbative WKB-active action is found to be (in the scaled $u\to gu$ perturbative form)
\begin{align}
    &\WP(u,g) = u+g^2 \left(\frac{25 u}{8}+\frac{5 u^3}{2}\right)+g^4
   \left(\frac{21777 u}{256}+\frac{5145 u^3}{32}+\frac{693
   u^5}{16}\right)\\
   \nonumber&+g^6 \left(\frac{12746305 u}{2048}+\frac{8703695
   u^3}{512}+\frac{1096095 u^5}{128}+\frac{36465 u^7}{32}\right)\\
   \nonumber &+g^8 \left(\frac{236185026649 u}{262144}+\frac{49058686105
   u^3}{16384}+\frac{16457464023 u^5}{8192}+\frac{470434965
   u^7}{1024}+\frac{37182145 u^9}{1024}\right)\\\nonumber
   &+\mathcal{O}(g^{10})
\end{align}
\subsubsection{Quintic Oscillator}
For the quintic oscillator, the quantum generators are defined  as 
\begin{equation}
    \mathcal{D}_n \coloneq \sum_{k=2(n-2)}^{2n-1} h_k^{(n)}(u)\frac{\del^k}{\del u^k}
\end{equation}
where for each $n$, we find the operators to be (up to $n=4$),
\begin{align}
   \mathcal{D}_2&= -\frac{1375 u}{81}-\frac{8750 u^2}{81}\del_{u}-\frac{11 \left(100000
   u^3-27\right)}{1512}\del_u^2-\frac{5}{567} \left(25000 u^4-27
   u\right)\del_u^3\\\nonumber
   \mathcal{D}_4&=\frac{2446767783-346260200000
   u^3}{267354138240}\del_u^4+\frac{15452613 u-45505100000
   u^4}{22558005414}\del_u^5\\&-\frac{25 \left(407675000 u^5+111051
   u^2\right)}{15038670276}\del_u^6-\frac{25525 \left(25000 u^6-27
   u^3\right)}{11279002707}\del_u^7\nonumber&
\end{align}
The perturbative WKB-active action is found to be (in the scaled $u\to gu$ perturbative
form)
\begin{align}
   &W_P(u,g) = u+g^3 \left(\frac{1107}{256}+\frac{1085 u^2}{32}+\frac{315
   u^4}{16}\right)\\\nonumber&+g^6 \left(\frac{118165905 u}{8192}+\frac{96201105
   u^3}{2048}+\frac{15570555 u^5}{512}+\frac{692835
   u^7}{128}\right)\\\nonumber&+g^9 \left(\frac{30926063193025
   u^4}{131072}+\frac{4140194663605 u^6}{32768}+\frac{456782651325
   u^8}{16384}+\frac{9704539845 u^{10}}{4096}\right)\\\nonumber
   &+\mathcal{O}(g^{12}).
\end{align}
\subsubsection{Octic Oscillator}
For the octic oscillator, the quantum generators are defined as 
\begin{equation}
    \mathcal{D}_n \coloneq \sum_{k = 3(n-2)}^{3n-1} h_k^{(n)}(u)\frac{\del ^k}{\del u^k}
\end{equation}
where for each $n$, we find the operators to be (up to n=4),
\begin{align}
   \mathcal{D}_2&=\left(\frac{17}{40}-\frac{2202112 u^3}{15795}\right)\del_u^{2}+\left(-\frac{5337088
   u^4}{15795}-\frac{7 u}{65}\right)\del_u^3+\left(\frac{8 u^2}{195}-\frac{163840
   u^5}{1053}\right)\del_u^4\\\nonumber&+\left(\frac{64 u^3}{585}-\frac{262144 u^6}{15795}\right)\del_u^5&
\end{align}
\begin{flalign}
    &\mathcal{D}_4= \scriptstyle\frac{337476819097141882388480
   u^7-1478909635385248358400 u^4-9845537056426404387
   u}{4169547442924907697705600}\del_u^8\nonumber &&
\end{flalign}
\begin{equation}
    \scriptstyle+\frac{1049040647461105660067840
   u^8+1034450914652600856576 u^5-8396644650494937129
   u^2}{32607163987248692229713325}\del_u^9\nonumber
\end{equation}
\begin{equation}
    \scriptstyle+\frac{96450605621156375429120
   u^9+146424199752518467584 u^6-438104226254951456
   u^3+1202959646624565}{25763685125727361761748800}\del_u^{10}\nonumber
\end{equation}
\begin{equation}
    \scriptstyle+\frac{26993024
   9942679613865984 u^{10}-1079196698132808204288
   u^7-6257546948502328320 u^4+10826636819621085
   u}{2086858495183916302701652800}\del_u^{11}\nonumber
\end{equation}
The perturbative WKB-active action is found to be (in the scaled $u\to gu$ perturbative
form)
\begin{align}
   & W_{\text{P}}(u,g) = u+g^3 \left(\frac{315}{128}+\frac{245 u^2}{16}+\frac{35
   u^4}{8}\right)\\\nonumber&+g^6 \left(\frac{5604849 u}{2048}+\frac{3209745
   u^3}{512}+\frac{291291 u^5}{128}+\frac{6435 u^7}{32}\right)\\\nonumber&+g^9
   \left(\frac{1030436816409 u^2}{131072}+\frac{151562676265
   u^4}{16384}+\frac{11952281341 u^6}{4096}+\frac{703227525
   u^8}{2048}+\frac{7436429 u^{10}}{512}\right)\\\nonumber
   &+\mathcal{O}(g^{12}).
\end{align}
\subsubsection{Septic Oscillator}
For the septic oscillator, the quantum generators are defined as 
\begin{equation}
    \mathcal{D}_n \coloneq \sum_{k = 3(n-2)}^{3n-1} h_k^{(n)}(u)\frac{\del ^k}{\del u^k}
\end{equation}
where for each $n$, we find the operators to be (up to n=2 because it gets long),
\begin{align}
   \mathcal{D}_2&=\frac{469957334 u^3}{2659375}-\frac{17802176084
   u^4}{2659375}\del_u+\left(\frac{25}{72}-\frac{29160010544 u
   ^5}{531875}\right)\del_u^2\\&-\left(\frac{7 u  \left(66357800768 u
   ^5+115625\right)}{9573750}\right)\del_u^3-\left(\frac{49 u ^2 \left(1133195168 u
   ^5-3125\right)}{4786875}\right)\del_u^4\\\nonumber&-\left(\frac{686 u ^3 \left(26353376 u
   ^5-3125\right)}{23934375}\right)\del_u^5&
\end{align}
The perturbative WKB-active action is found to be (in the scaled $u\to gu$ perturbative
form)
\begin{align}
    W_{\text{P}}(u,g) &= u+\left(\frac{180675}{2048}+\frac{444381 u^2}{512}+\frac{82005
   u^4}{128}+\frac{3003 u^6}{32}\right)g^5\\\nonumber&+\Bigg(\frac{182627818702875 u}{2097152}+\frac{156916927352185
   u^3}{524288}+\frac{13513312267455 u^5}{65536}\\\nonumber &+\frac{824707412529
   u^7}{16384}+\frac{43689020375 u^9}{8192}+\frac{456326325
   u^{11}}{2048}\Bigg) g^{10}+O\left(g^{15}\right).
\end{align}
\section{Nonperturbative action integral and determination of the matrix $M_*$}\label{app:WNP}

Here, we present how to determine the correct linear combination for nonperturbative actions $W_{\text{NP}}$, i.e., the diagonal scaling matrix $S$ and the $\text{SL}(2\text{g},\mathbb{Z})$ transformation $M_0$ giving the choice for the product matrix $M_*$. For demonstration, we will perform the calculations explicitly for the sextic oscillator.

\subsection{Direct integration of the nonperturbative action integral}
In order to fix the linear combination coefficients of the nonperturbative action $W_{\text{NP}}$ in terms of the PF-basis, we need to integrate the action integral directly. To our advantage, we only need to compute the first $2\text{g}$ orders of the classical action. 

The nonperturbative classical action is given by 
\begin{equation}
    W_{\text{NP}}^{(0)}(u) = -i\oint_{\gamma_{\text{NP}}} pdq = -2i\int_{q_1}^{q_2}\sqrt{2(u-V(q))}dq 
\end{equation}
where $\gamma_{\text{NP}}$ is the cycle encircling the turning points $q_1$ and $q_2$ whose interval $I=[q_1(u),q_2(u)]$ defines the nonperturbative, classically forbidden, region at energy parameter u. 
The boundaries of the interval $I=[q_1(0),q_2(0)]$ are the zeroes of $V(q) = 0$ and therefore the branch points of the square root. Notice that at exactly $u=0$, the turning point $q_1$ coincides with the other turning point $-q_1$ at the critical point $q=0$ which makes the integral singular. We have to treat the singular behavior of the function at this point separately. In order to perform the integration, we split the interval into three regions. 
\begin{equation}
    I= \underbrace{(q_1,q_1+\epsilon_1]}_{= I_1}\cup\underbrace{[q_1+\epsilon_1,q_2-\epsilon_2]}_{= I_2}\cup \underbrace{[q_2-\epsilon_2, q_2]}_{= I_3}
\end{equation}
After integration the limit $\epsilon_1,\epsilon_2 \to 0$ is to be taken. As expected, the most relevant finite contribution comes from the interval $I_2$. We will compute the integral over this middle region by making an expansion in $u$ around the singular point $u\sim 0$,
\begin{equation}
    -2i\int_{I_2}\sqrt{2(u-V(q))}dq= 2\sqrt{2}\sum_{k=0}^{\infty}(-1)^k\begin{pmatrix}1/2\\ k
    \end{pmatrix}u^k\int_{q_1+\epsilon_1}^{q_2-\epsilon_2} dq [V(q)]^{\frac{1}{2}-k}.
\end{equation}
Direct integration of the above expansion gives the leading orders
\begin{align}
    -2i\int_{I_2}\sqrt{2u-q^2+2q^6}dq =  \frac{\pi}{4\sqrt{2}} +u\left(2\log(\epsilon_1)-\frac{1}{2}\log2\right) - u^2\frac{1}{2\epsilon_1^2} \\\nonumber+ u^3\left(\frac{31}{12}-\frac{5}{4}\log 2 - 5\log(\epsilon_1) -\frac{1}{4\epsilon_1^4}\right) + \mathcal{O}(u^4)
\end{align}
where we have omitted powers of $\epsilon_1, \epsilon_2$ since they will vanish, and also omitted singular terms that depend on $\epsilon_2$ because they are exactly canceled by the integral over the region $I_3$. Since the turning point $q_2$ is always a regular point as $u\to 0$, the contribution of the integral over region 3 only results in the cancellation of the singular terms due to $\epsilon_2$ and doesn't give any finite result. 
For the integral over region $I_1$, we make a change of variable $v^2=2V(q)$, 
\begin{equation}
    -2i\int_{I_1}\sqrt{2u-q^2+2q^6}dq = 2\int_{\sqrt{2u}}^{V(q_1+\epsilon_1)}(v^2 - 2u)^{1/2}\left(\frac{d q}{d v}\right)dv
\end{equation}
where one can find $q(v)$ by inverting the series $v(q) = q(1-2q^4)^{1/2} = q-q^5+\frac{1}{2}q^9+\cdots$. After integrating it term by term and expanding around $\epsilon_1\sim 0$, one gets
\begin{align}
    -2i\int_{I_1}\sqrt{2u-q^2+2q^6} =  u\Big(-1 + \log u - \log2 - 2\log(\epsilon_1)\Big) + u^2\frac{1}{2\epsilon_1^2}\\\nonumber+u^3\left(\frac{37}{12}-\frac{5}{2}\log2 +\frac{5}{2}\log u +5\log(\epsilon_1) \right) + \mathcal{O}(u^4).
\end{align}
\subsection{Linear combination of the WKB-active cycles}
Finally, adding up the results of each region, we get the classical action of the WKB-active cycle for the sextic potential
\begin{equation}
    W_{\text{NP}}(u)= \frac{\pi}{4\sqrt{2}} + u\left(-1+\log\left(\frac{u}{2^{3/2}}\right)\right) + u^3\left(\frac{17}{3}+\frac{5}{2}\log\left(\frac{u}{2^{3/2}}\right)\right)+\mathcal{O}(u^5)
\end{equation}
Given that one has the PF-basis as in \eqref{sexticPF}, we identify the linear combination to be
\begin{equation}
    W_{\text{NP}}(u) = a_1^D(u) + \frac{\pi}{4\sqrt{2}}a_2^D(u) 
\end{equation} 
which extends to the quantum action $W_{\text{NP}}(u,g)$ as well. This is sufficient information for choosing the matrix $M_*$ to be as given in the main text, i.e.,
\begin{equation}
     S=\text{diag}\left(1,1,1,\frac{\pi}{4\sqrt{2}}\right),\quad M_o = \begin{pmatrix}
         1 & 0 &0 &0\\
         0 & 1 &0 &0\\
         0 & 0 &1 &1\\
         0 & 0 &0 &1
     \end{pmatrix} \implies M_oS = M_* = \begin{pmatrix}
         1 & 0 &0 &0\\
         0 & 1 &0 &0\\
         0 & 0 &1 &\frac{\pi}{4\sqrt{2}}\\
         0 & 0 &0 &\frac{\pi}{4\sqrt{2}}
     \end{pmatrix}
 \end{equation}
which is used to derive the modified symplectic intersection matrix via $M_*^TK'M_*=K$.

\subsection{Octic oscillator}
For the octic oscillator the expansion of the barrier action $W^{(0)}_{\text{NP}}(u)=-2i\int_{q_1}^{q_2}\sqrt{2(u-V(q))}\,dq=2\int_{q_1}^{q_2}\sqrt{2V(q)-2u}\,dq$ can be done in closed form. Expanding the integrand in $u$ and continuing the resulting Beta integrals analytically (substitute $t=2q^6$), the coefficient of $u^k$ is
\begin{equation}
    w_k=\begin{pmatrix}1/2\\ k\end{pmatrix}(-2)^k\,\frac{1}{3}\,2^{(k-1)/3}\,B\Big(\frac{1-k}{3},\frac{3}{2}-k\Big),\qquad k\not\equiv1\ (\text{mod }3).
\end{equation}
For $k\equiv1$ the Beta function has a pole: these are the powers $u^{1+3m}$ that come with $\log u$ and are reproduced by $a_1^D$. The remaining powers are matched by $c_2^Da_2^D$ ($k\equiv0$) and by $c_2a_2$ ($k\equiv2$), which gives $c_2^D=w_0=\frac{1}{3}\,2^{-1/3}B(\frac{1}{3},\frac{3}{2})=S_0$ and $c_2=w_2$; with $\Gamma(\frac16)=2^{-1/3}\sqrt{3/\pi}\,\Gamma^2(\frac13)$ these are the values quoted in Sec.~\ref{sec:octic}. All higher coefficients are then checks, for example $w_3/w_0=-\frac{455}{324}$ and $w_5/w_2=\frac{124729}{6480}$, and the term linear in $u$, $-(1+\log2^{4/3})\,u$, fixes the constant in the logarithm of $a_1^D$. We have also checked the result against high-precision numerical integration of $W^{(0)}_{\text{NP}}(u)$.
\section{Vanishing of the residue term for anharmonic
oscillators}\label{app:residue}
In Sec.~\ref{sec:examples} we used that for $V(q)=\frac{1}{2}q^2-q^d$
with $d\geq5$ the right-hand side of the bilinear identity
\eqref{quantumPNP} vanishes. The reason is simple: the residue is a
sum over the saddles of $V$, and for these potentials the instantons
differ only by phases that are roots of unity.

\paragraph{The residue as a sum over saddles.} Let us start with the
classical relation \eqref{classicalPNP1}, whose right-hand side is
\begin{equation}
    \mathcal{R}_0\equiv\sum_{q_0\in\text{poles}}\text{Res}\left(f\,\frac{dq}{p},q_0\right),\qquad
df=p\,dq .
\end{equation}
Since $p\,dq$ and $dq/p$ are regular at the turning points, only
$q=\infty$ contributes (one point for odd $d$, two for even $d$). For
a general polynomial
$V(q)=v_dq^d+v_{d-1}q^{d-1}+v_{d-2}q^{d-2}+\dots$, a short large-$q$
expansion gives
\begin{equation}
    \mathcal{R}_0=-\frac{8}{d^2-4}\left(\frac{v_{d-2}}{v_d}-\frac{d-1}{2d}\,\frac{v_{d-1}^2}{v_d^2}\right)=-\frac{4d^2}{d^2-4}\sum_{V'(q_c)=0}\frac{V(q_c)}{V''(q_c)},
    \label{eq:saddlesum}
\end{equation}
where the second form follows from the residue theorem applied to
$V\,dq/V'$. Notice that only the two subleading coefficients of $V$
enter, in the combination that is invariant under shifts of $q$. This
is dictated by the scaling $q\to\lambda q$, under which
$\mathcal{R}_0$ has weight $2$ while $u$ has weight $d$. In words,
every saddle contributes its height over its curvature. For the
multi-well potentials \eqref{eq:multiwell} the minima have $V=0$,
every barrier top contributes a positive amount, and
\eqref{eq:saddlesum} reproduces the constant $\mathcal{S}$ of
\eqref{eq:mwmain}.

\paragraph{Roots of unity.} For $V=\frac{1}{2}q^2-q^d$ the minimum
$q=0$ has $V=0$ and drops out. The remaining saddles form the single
orbit $q_k=d^{-1/m}\zeta^k$ of $R$ (Fig.~\ref{fig:sym}), with heights
$V(q_k)=u_*\zeta^{2k}$, $u_*=\frac{m}{2d}\,d^{-2/m}$, and the same
curvature $V''(q_k)=-m$ for all $k$. Therefore
\begin{equation}
    \mathcal{R}_0=\frac{4d^2}{d^2-4}\,\frac{u_*}{m}\,\sum_{k=0}^{m-1}\zeta^{2k}.
    \label{eq:rootsofunity}
\end{equation}
The instantons enter only through their phases $\zeta^{2k}$, the same
phases that relate their actions, $S_k=\zeta^{2k}S_0$. For $m\geq3$,
i.e.\ $d\geq5$, $\zeta^2\neq1$ is an $m$-th root of unity and the sum
vanishes: the $m$ instantons cancel each other in the residue. For
$m=1,2$ (cubic and quartic) $\zeta^2=1$ and they add up instead,
giving $\mathcal{R}_0=\frac{2}{15}$ and $\frac{1}{3}$. Interestingly,
these are exactly the instanton actions $S_0$ of $\frac{1}{2}q^2-q^3$
and $\frac{1}{2}q^2-q^4$, the constants of the genus-one P/NP relation
\cite{Dunne:2014bca,Basar:2017hpr}. The two classes of potentials
therefore differ only in how their instantons add up: with the same
sign for multi-wells, and as the $m$-th roots of unity for the
anharmonic oscillators.

\paragraph{No quantum corrections.} Let us keep the coefficient of the
quadratic term as a parameter, $V_\mu(q)=\frac{1}{2}\mu q^2-q^d$. Then
$R$ extends to a one-parameter group of scalings,
\begin{equation}
    (q,\,p,\,u,\,\mu,\,g)\longmapsto(\lambda
q,\,\lambda^{d/2}p,\,\lambda^{d}u,\,\lambda^{d-2}\mu,\,\lambda^{d/2+1}g),\qquad
\lambda\in\mathbb{C}^{\times},
    \label{eq:scaling}
\end{equation}
which leaves the curve and the Riccati equation invariant; for
$\lambda=\zeta$ it reduces to $R$, up to $(p,g)\to(-p,-g)$. The
quantum actions $\frac{1}{g}\oint p_Q\,dq$ are invariant, so the
coefficient of $g^{2n-3}$ on the right-hand side of \eqref{quantumPNP}
is a sum of monomials $u^j\mu^k$ with
\begin{equation}
    dj+(d-2)k=2-n(d+2).
    \label{eq:weight}
\end{equation}
For $n\geq1$ the right-hand side of \eqref{eq:weight} is negative.
Hence the residue term receives no quantum corrections,\footnote{This
is immediate for the residue at $q=\infty$, whose coefficients are
polynomials in $u$ and $\mu$. The quantum corrections to $p_Q$ also
have poles at the turning points; summed over the turning points they
give a rational function of $u$ whose only possible poles are at the
critical values $u_k$. These poles are absent, since near $u=u_k$ the
singular part of the dual action depends on $u$ and $g$ only through
the vanishing action $B$, as $F(B)+c\,B\log g$ (cf.~\eqref{logform}),
and its bracket with $B$ is regular. We have also verified the
vanishing explicitly for the octic through order $g^{15}$.} and it
equals $-\mathcal{R}_0/g^3$, which vanishes for $d\geq5$. (For $n=0$,
\eqref{eq:weight} allows only $\mu^2$ for $d=3$ and $\mu$ for $d=4$,
in agreement with \eqref{eq:rootsofunity}.)

\paragraph{The sextic.} The same counting applied to $p\,dq$ itself,
of weight $\frac{d}{2}+1$, allows a residue at infinity only if
$dj+(d-2)k=\frac{d}{2}+1$. Among $d\geq3$ this happens only for $d=6$,
where $\text{Res}_{\infty_\pm}p\,dq\propto\mu$ is independent of $u$.
Then $f$ has a logarithm at $\infty_\pm$, and the right-hand side
picks up, besides \eqref{eq:saddlesum}, a term proportional to the
Abel-Jacobi integral $\int_{\infty_-}^{\infty_+}dq/p$. This residue is
the constant solution $a^{(0)}_{2,D}=1$ of \eqref{pfsextic}, and its
partner $a_2$ is the corresponding Abel-Jacobi integral. With this
pair included in $\Pi$, as in Sec.~\ref{sec:examples}, it accounts for
the extra term, and the remaining residue is \eqref{eq:rootsofunity}
with $\zeta^2=-1$, which vanishes.

\bibliographystyle{utphys}
\bibliography{GutzSumRef}
\end{document}